\documentclass[prd,twocolumn,superscriptaddress,floatfix,nopacs,preprintnumbers,nofootinbib]{revtex4}
\usepackage[utf8]{inputenc}

\usepackage{graphicx}
\usepackage[normalem]{ulem}
\usepackage{xcolor}
\usepackage{adjustbox}
\usepackage{mathrsfs}
\usepackage{amssymb,bm}
\usepackage{amsmath}
\usepackage{mathtools}
\usepackage{physics}
\usepackage{slashed}
\usepackage{verbatim}
\usepackage{ragged2e}
\usepackage{caption}
\usepackage{subcaption}
\usepackage{multirow}
\usepackage{array}
\usepackage{tikz}
\usepackage{lipsum} 
\usepackage{soul}
\newcommand{\nc}{N_\mathrm{c}}
\newcommand{\aem}{\alpha_\mathrm{em}}
\newcommand{\gev}{\mathrm{GeV}}

\newcommand{\bt}{\mathbf{b}}
\newcommand{\qso}{Q_{\mathrm{s},0}} % Q_{0} or Q_{\mathrm{s},0}
 
\newcommand{\rt}{\mathbf{r}}

\newcommand{\kt}{\mathbf{k}}

\newcommand{\xt}{\mathbf{x}}

\newcommand{\lqcd}{\Lambda_\text{QCD}}
\newcommand{\cf}{C_\mathrm{F}}

\newcommand{\xij}[1]{\mathbf{x}_{#1}}

\newcommand{\btheta}{\boldsymbol{\theta}}
\newcommand{\ymodel}{\mathbf{y}(\btheta)}

\newcommand{\yexp}{\mathbf{y}_{\mathrm{exp}}}
\newcommand{\deltay}{\Delta \mathbf{y}(\btheta)}

\newcommand{\sigmaltnlo}{\sigma_{L,T}^{\textrm{NLO}}}
\newcommand{\sigmaltdip}{\sigma_{L,T}^{\textrm{dip}}}

\newcommand{\sigmaltqgu}{\sigma_{L,T}^{qg, \textrm{unsub.}}}
\newcommand{\sigmalt}[1]{\sigma_{L,T}^{#1}}

\newcommand{\g}{\gamma}

\newcommand{\csq}{C^2}
\newcommand{\mc}{m_c}
\newcommand{\initsig}{\sigma_0/2}

\newcommand{\as}{\alpha_\mathrm{s}}

\usepackage[breaklinks,colorlinks,citecolor=citcolor,urlcolor=blue,linkcolor=lcolor]{hyperref}
\definecolor{lcolor}{rgb}{0.5,0,0}
\definecolor{citcolor}{rgb}{0,0.3,0.0}
\definecolor{teal}{rgb}{0.0, 0.5, 0.5}

\begin{document}

\title{Initial condition for the Balitsky-Kovchegov equation at next-to-leading order }

\author{Carlisle Casuga}
\email{carlisle.doc.casuga@jyu.fi}
\affiliation{
Department of Physics, University of Jyväskylä,  P.O. Box 35, 40014 University of Jyväskylä, Finland
}
\affiliation{
Helsinki Institute of Physics, P.O. Box 64, 00014 University of Helsinki, Finland
}
\author{Henri Hänninen}
\email{henri.j.hanninen@jyu.fi}
\affiliation{Department of Mathematics and Statistics, University of Jyväskylä, P.O. Box 35, 40014 University of Jyväskylä, Finland}
\author{Heikki Mäntysaari}
\email{heikki.mantysaari@jyu.fi}
\affiliation{
Department of Physics, University of Jyväskylä,  P.O. Box 35, 40014 University of Jyväskylä, Finland
}
\affiliation{
Helsinki Institute of Physics, P.O. Box 64, 00014 University of Helsinki, Finland
}

\begin{abstract}

    We determine the initial condition of the Balitsky-Kovchegov (BK) evolution equation at next-to-leading order (NLO) accuracy using HERA deep inelastic scattering data. We use Bayesian inference to compare model values, derived from NLO impact factors and the BK evolution equation with kinematical constraint resumming most important higher order corrections, to precise data and to extract posterior distributions characterizing the initial condition. The total cross section and charm quark production data from HERA are found to provide stringent constraints on the posterior. These distributions quantify the uncertainty in the initial condition and serve as necessary input for propagating uncertainties to all NLO calculations within the Color Glass Condensate framework.
    
\end{abstract}

\maketitle

\section{Introduction}

The Color Glass Condensate (CGC)~\cite{Iancu:2003xm}  effective field theory provides a natural framework for describing the gluon-dense, high-energy structure of  protons and nuclei. As the collision energy increases or the Bjorken-$x$ decreases at fixed probe virtuality $Q^2$, one unlocks the saturation regime with emergent non-linear QCD effects. This regime is characterized by the saturation scale, $Q_s^2$, which scales as $Q_s^2 \sim A^{1/3}$, where $A$ is the nuclear mass number. In this context, the upcoming Electron Ion collider (EIC)~\cite{AbdulKhalek:2021gbh} (or other potential future nuclear-DIS facilities~\cite{LHeC:2020van,Anderle:2021wcy}) will feature heavy nuclear targets and provide access to the saturated state of gluonic matter. 

In the leading order (LO) dipole picture of deep inelastic scattering (DIS), the incoming virtual photon splits into a quark-antiquark pair which interacts with the target nucleus~\cite{Mueller:1994jq}. 
The necessary ingredients in calculations of DIS structure functions in the CGC framework are the DIS impact factor and the small-$x$ Balitsky-Kovchegov (BK)~\cite{Kovchegov:1999yj,Balitsky:1995ub} evolution equation at the desired order in perturbation theory.

First principle calculations of impact factors at NLO in $\as$ have been performed in the light cone perturbation theory for light~\cite{Balitsky:2010ze, Balitsky:2012bs, Hanninen:2017ddy, Beuf:2017bpd, Beuf:2016wdz, Beuf:2011xd} and massive~\cite{Beuf:2022ndu,Beuf:2021qqa,Beuf:2021srj} quarks. These impact factors, which describe the photon splitting into partonic states, are coupled to the high-energy BK evolution equation that resums contributions enhanced by large logarithms of energy  $\sim \as \ln(1/x)$ (or $\as^2 \ln(1/x)$ at NLO). The full NLO BK evolution equation is derived in Ref. \cite{Balitsky:2007feb} with numerical solutions implemented in Refs. \cite{Lappi:2015fma, Lappi:2016fmu}. Prescriptions that resum dominant single and double logarithmic contributions, that provide an accurate approximation to the NLO BK equation~\cite{Lappi:2016fmu}, have been obtained in Refs.~\cite{Ducloue:2019ezk, Iancu:2015vea, Iancu:2015joa, Beuf:2014uia}. For this study, we adopt the BK evolution equation that resums the large double logarithmic terms to approximate the full NLO BK equation. Finite-$N_c$ corrections to the NLO BK equation, found to have a small effect on dipole amplitudes \cite{Lappi:2020srm}, and other  corrections like the resummation of single logarithmic terms and higher order terms not enhanced by transverse logarithms, are omitted from the present analysis. Furthermore, we do not include  dependence on the impact parameter, as studied in Ref. \cite{Cepila:2024qge}.

The BK equation describes the energy dependence of the 
dipole-target scattering amplitude. This perturbative evolution equation takes as a non-perturbative input the dipole-target scattering amplitude at an initial energy scale, and predicts its energy evolution towards higher energies. Models like the McLerran-Venugopalan (MV)~\cite{McLerran:1993ni} model or the Golec-Biernat-W{\"u}sthoff (GBW) model~\cite{GolecBiernat:1998js} provide reasonable parametrizations for this initial condition. Substantial efforts have been made in extracting the initial condition from available data at LO \cite{Albacete:2009fh, Albacete:2010sy, Lappi:2013zma}, and at NLO with light~\cite{Beuf:2020dxl} and heavy~\cite{Hanninen:2022gje} quarks. In particular, Ref.~\cite{Hanninen:2022gje} shows the feasibility of a global fit at NLO accuracy by demonstrating that it is possible to simultaneously describe both the total cross section~\cite{H1:2015ubc} and heavy quark production~\cite{H1:2018flt} data from HERA at small $x$. These previous NLO dipole fits, however, do not include the heavy quark contribution in the actual fit procedure, and neglect correlated experimental uncertainties. Furthermore, they also  lack a systematic quantification of the uncertainty of the determined non-perturbative BK initial condition, which has been obtained in a leading order analysis in Ref.~\cite{Casuga:2023dcf}. 
An accurate extraction of this non-perturbative input, among with well-defined uncertainty estimates, is  essential in order to promote the precision in theoretical calculations to the level that is comparable to current and future  high-precision measurement sensitive to saturation effects at the EIC or at the LHC~\cite{Morreale:2021pnn}.

The main objective of this work is to extend our previous analysis~\cite{Casuga:2023dcf} to infer the initial condition for the BK equation in NLO accuracy, combining NLO impact factors with dipole amplitudes evolved using an evolution equation of partial NLO accuracy to compute the DIS structure functions. For the first time, we perform a global analysis including both the HERA total cross section and the charm production data as constraints. To extract the model parameters in the initial condition, we use Bayesian inference, which not only identifies the best-fit values but also probes its neighboring area in the parameter space. This approach determines posterior distributions that provide a streamlined method to propagate the initial condition uncertainty to calculations of other observables within the CGC framework at NLO. 

The paper is structured as follows. Sec.~\ref{sec:dis} provides a brief summary of the DIS cross section calculation at the next-to-leading order accuracy in the CGC approach. This involves a two-part discussion of the NLO impact factors in Sec.~\ref{subsec:dipolepicture} and the approximated NLO evolution equation in Sec.~\ref{subsec:evolution}. Details of the Bayesian inference setup used to find the probability distribution describing the initial condition to the BK evolution equation are presented in Sec.~\ref{sec:bayesian}. The obtained posterior distributions, best-fit values, and other results are then shown in Sec.~\ref{sec:results}. Conclusions and main findings are summarized in Sec.~\ref{sec:conclusions}.

\section{Deep inelastic scattering at NLO in dipole picture}
\label{sec:dis}
\subsection{NLO cross section}
\label{subsec:dipolepicture}

The leading order dipole picture of DIS consists of a photon fluctuating to a dipole and interacting instantaneously with the target hadron by exchanging low-$x$ gluons. The separation between the
lifetime of the dipole and the timescale of the interaction enables one to 
factorize the leading order $\gamma^*A$ cross section, as a function of Bjorken-$x$ and photon virtuality, $Q^2$, as~\cite{Kovchegov:2012mbw}
\begin{multline}
\label{eq:lo-cross-section}
    \sigma^{\gamma^* p}_{T,L}(x,Q^2)=2 \sum_f \int \dd[2]{\bt} \dd[2]{\rt} \dd{z} |\psi^{\gamma^* \to q\bar q}_{T,L}(\rt,Q^2,z)|^2
    \\
    \times
    N_{\textrm{LO}}(\rt,\bt,x).
\end{multline}
The light-cone wave function, $\psi^{\gamma^*\rightarrow q\bar{q}}_{T,L}$, describes the splitting of the virtual photon to the quark-antiquark pair with a transverse separation $\rt$, where the quark carries a fraction, $z$, of the large photon plus momentum. Here $T$ and $L$ refer to the transverse and longitudinal photon polarization, respectively. The dipole interacts with the target color field  eikonally, so that $\rt$ does not change during the interaction. In this work we further neglect the impact parameter dependence from the dipole-target scattering amplitude $N_{\textrm{LO}}$ and, following e.g. Refs.~\cite{Albacete:2010sy,Lappi:2013zma}, replace $\int \dd[2]{\bt} \to \initsig$ where $\initsig$ is the proton transverse area and $\bt$ is the impact parameter in the dipole-target scattering.

The total DIS cross section data is typically reported as a reduced cross section defined as
\begin{equation}
    \sigma_r(y,x,Q^2) = F_2\left(x,Q^2\right) - \frac{y^2}{1+(1-y)^2} F_L\left(x,Q^2\right).
\end{equation}
Here the inelasticity is defined as $y = Q^2/(sx)$, where $\sqrt{s}$ is the center-of-mass energy in the electron-proton scattering.
 The structure functions $F_2$ and $F_L$ are related to the virtual photon-proton cross section as
\begin{align}
    F_2 &= \frac{Q^2}{4\pi^2 \aem} (\sigma^{\gamma^* A}_T + \sigma^{\gamma^* A}_L), \\
    F_L &= \frac{Q^2}{4\pi^2 \aem} \sigma^{\gamma^* A}_L.
\label{eq:f2fl}
\end{align}

The NLO corrections to the $\sigma^{\gamma^* A}_{L, T}$ cross section include the tree-level contribution describing the photon fluctuating to a $q\bar{q}g$ system, where the additional gluon also  interacts with the target. Furthermore, NLO corrections also include virtual one-loop corrections to the light-cone wave function describing the $\gamma^{*} \rightarrow q \bar{q}$ splitting. The $q\bar{q}g$ and the one-loop $q\bar{q}$ contributions in the dipole factorization have been first calculated in the massless limit in Refs.~\cite{Hanninen:2017ddy,Beuf:2016wdz,Beuf:2017bpd}, and later results with a finite quark mass enabling also calculations of heavy quark production were published in Refs.~\cite{Beuf:2022ndu,Beuf:2021srj,Beuf:2021qqa}. 

At NLO accuracy, a rapidity divergence appears when the gluon longitudinal momentum fraction $z_2\to 0$. This divergence  needs to be absorbed into the Balitsky-Kovchegov evolution of the dipole-target scattering amplitude, and there are different scheme choices that can be employed. In this work, we apply the ``unsubtracted'' scheme following Refs.~\cite{Iancu:2016vyg,Beuf:2017bpd,Ducloue:2017ftk}, which has been previously applied in small-$x$ phenomenology e.g. in Refs.~\cite{Beuf:2020dxl,Mantysaari:2021ryb,Hanninen:2022gje}. In the unsubtracted scheme the gluon longitudinal momentum  fraction $z_2$ is limited from below by requiring that the invariant mass of the $q\bar q g$ system cannot exceed the center-of-mass energy of the photon-nucleon system $W$. 

In the unsubtracted scheme, the NLO cross section for the $\gamma^*$-target scattering can be written as~\cite{Beuf:2020dxl}
\begin{equation}
    \sigmaltnlo = \sigmalt{\textrm{IC}} + \sigmaltdip + \sigmaltqgu .
\end{equation}
Here  $\sigmalt{\textrm{IC}}$ is the lowest order contribution from the $q\bar q$-target scattering without including the BK evolution, $\sigmaltqgu$ describes the contribution where the $q\bar{q}g$ system interacts with the target, and the dipole contribution $\sigmaltdip$ contains loop corrections to the $q\bar q$-target scattering. Note that when the cross section is written in this form, UV divergences have been cancelled between $\sigmaltdip$ and $\sigmaltqgu$ and as such this division is not unique.

The lowest order contribution $\sigmalt{\textrm{IC}}$ can be evaluated using Eq.~\eqref{eq:lo-cross-section} together with the dipole amplitude evaluated at the initial condition of the BK evolution. This is different from the leading order dipole picture result, as in the small-$x$ power counting  $\as \ln 1/x \sim 1$, and terms of this form are resummed by having an $x$-dependent dipole amplitude satisfying the BK equation. In the unsubstracted scheme applied here, the BK evolution dynamics is included in the term $\sigmaltqgu$. 

The term describing the $q\bar q g$-target interaction can be written as
\begin{equation}
    \sigmaltqgu = K_{q\bar qg} \otimes N_{012}. 
    \label{eq:sigma_qqg} 
\end{equation}
Similarly for the $q\bar q$-target scattering at NLO we have
\begin{equation}
\label{eq:sigma_qq}
    \sigmaltdip = K_{q\bar q} \otimes N_{01}.
\end{equation}
The convolutions in Eqs. \eqref{eq:sigma_qqg} and \eqref{eq:sigma_qq} are over the transverse coordinates $\xt_i$ and the parton longitudinal momentum fractions $z_i$.
Explicit expressions for the hard factors $K_{q\bar q g}$ and $K_{q\bar q}$ with non-zero quark masses employed in this work can be found from Refs.~\cite{Beuf:2022ndu,Beuf:2021srj,Beuf:2021qqa}, for brevity we do not repeat these expressions in this work. 

Eikonal quark(gluon)-target interaction is described in terms of Wilson lines in fundamental (adjoint) representation. The amplitude for the quark-antiquark dipole to scatter off the dense CGC target is $N_{01}=1-S_{01}$, where
\begin{equation}
     S_{01} = \frac{1}{\nc} \left\langle  \Tr{V(\xt_0) V^\dagger(\xt_1)} \right\rangle . \label{eq:s01}
\end{equation}
The average $\langle \cdot \rangle$ refers to an average over the target color field configurations.
Here $\xt_0$ and $\xt_1$ are the quark and antiquark transverse coordinates, and $N_{01} = N(|\xij{01}|)$ with $\xt_{ij} = \xt_i - \xt_j$. As mentioned above, the impact parameter dependence of the dipole amplitude $N_{01}$ is neglected in this work (see also Refs.~\cite{Bendova:2019psy,Mantysaari:2024zxq,Cepila:2024qge} for  recent discussions of the impact parameter dependence in the BK evolution). Similarly, the scattering amplitude for the $q\bar q g$ system $N_{012} = 1-S_{012}$ is~\cite{Hanninen:2017ddy}
\begin{equation}
    S_{012} = \frac{\nc}{2\cf} \left( S_{02}S_{21} - \frac{1}{\nc^2} S_{01}\right). \label{eq:s012}
\end{equation}

In the unsubtracted scheme, as already mentioned above, one introduces a lower limit for the gluon longitudinal momentum fraction $z_2$ when evaluating Eq.~\eqref{eq:sigma_qqg}. Following Ref.~\cite{Beuf:2020dxl}, this limit is 
\begin{equation}
    z_2 > z_{2,\mathrm{min}} = \frac{Q_0^2}{W^2}.
    \end{equation}
Again following \cite{Beuf:2020dxl} we set the non-perturbative target momentum scale $Q_0^2=1\,\mathrm{GeV}^2$. 

When the gluon loop corrections to the $q\bar q$-target scattering, Eq.~\eqref{eq:sigma_qq}, are calculated in Refs.~\cite{Beuf:2022ndu,Beuf:2021srj,Beuf:2021qqa,Beuf:2016wdz}, the integral over $z_2$ has been  calculated analytically, integrating down to $z_2=0$. This is possible as, unlike when evaluating $\sigmaltqgu$, the $z_2$ integral in $\sigmaltdip$ is not logarithmically divergent and the contribution from $z_2 < z_{2,\mathrm{min}}$ can be expected to be small.  

\subsection{Approximative BK evolution at next-to-leading order}
\label{subsec:evolution}
The dipole-target scattering amplitude depends on the energy (or Bjorken-$x$), and in the large-$\nc$ limit this evolution is described in terms of the BK equation. The BK equation at next-to-leading order has been derived in Ref.~\cite{Balitsky:2007feb}, and the most important higher-order contributions enhanced by large transverse logarithms have been resummed to all orders using different prescriptions in Refs.~\cite{Beuf:2014uia,Iancu:2015joa,Iancu:2015vea,Ducloue:2019ezk}.  

In this work, we approximate the next-to-leading order BK equation by the ``Kinematically Constrained BK equation'' (KCBK) from Ref.~\cite{Beuf:2014uia} (following the terminology of Ref.~\cite{Beuf:2020dxl}). The KCBK equation resums the most important contributions enhanced by double transverse logarithms, which in practice is obtained by imposing the required time ordering between the consecutive gluon emissions in the evolution. It has been shown in Ref.~\cite{Lappi:2016fmu} that when such higher order contributions are resummed to the leading order equation, a good approximation for the NLO BK equation is obtained. 
We have also explored a setup that additionally resums contributions enhanced by single transverse logarithms~\cite{Iancu:2015joa}, 
but such setup did not result in good fits as we will discuss in more detail in Sec.~\ref{sec:results}.
The inclusion of numerically more demanding NLO corrections not enhanced by transverse logarithms to the evolution is left for future work. In that case the resummation of single transverse logarithms will also be  crucial~\cite{Iancu:2015joa}). 

The KCBK evolution equation reads
\begin{align}
\label{eq:kcbk}
    & \partial_Y S_{01}(Y) =
    \notag
    \\
    & \, \int \dd[2]{\xij{2}}
    K_\text{BK}(\xij{0}, \xij{1}, \xij{2})
        \theta\left(Y - \Delta_{012} \right)
    \notag
    \\
    & \, \, \times \! \left[ S_{02}(Y \! - \! \Delta_{012}) S_{21}( Y \! -\! \Delta_{012})- S_{01}(Y)  \right].
\end{align}
Here the evolution rapidity $Y$ is defined using the gluon plus momentum $k^+ = z_2 q^+$ where $q^+$ is the large photon plus momentum:
\begin{equation}
    Y = \ln \frac{k^+}{P^+}.
\end{equation}
Here $P^+ = Q_0^2/(2P^-)$ is the target plus momentum. 

In this work we apply two different running coupling prescriptions. First, the coordinate space strong coupling constant in the BK evolution and in the impact factor can be evaluated at the distance scale set by the parent dipole $|\xt_{01}|$. In this case the BK kernel, which describes the probability density to emit a gluon to the transverse position $\xt_2$, reads
\begin{equation}
     K_\text{BK}(\xij{0},\xij{1}, \xij{2}) = \frac{\nc \as(\xij{01}^2)}{2\pi^2} 
        \frac{\xij{01}^2}{\xij{12}^2 \xij{02}^2}
\end{equation}
We refer to this as the parent dipole prescription. The other option, referred to as the Balitsky+smallest dipole prescription ~\cite{Balitsky:2006wa}, corresponds to the BK kernel
\begin{multline}
    \label{eq:bk-rc-balitsky}
  K_\text{BK}(\xij{0},\xij{1}, \xij{2}) = \frac{\nc \as(\xij{01}^2)}{2\pi^2} \left[
        \frac{\xij{01}^2}{\xij{12}^2 \xij{02}^2} \right. \\
        + \frac{1}{\xij{02}^2} \left( \frac{\as(\xij{02}^2)}{\as(\xij{12}^2)} -1 \right) 
        + \left. \frac{1}{\xij{12}^2} \left( \frac{\as(\xij{12}^2)}{\as(\xij{02}^2)} -1 \right)
  \right].
\end{multline}
In the limit where one of the daughter dipoles $\xij{02},\xij{12}$ is very small, the running coupling scale is set by the smallest dipole, and the Balitsky prescription typically results in slower evolution compared to the parent dipole one. As such, in this prescription we evaluate the running coupling in the impact factor at the scale set by the smallest dipole size:  $\as = \as(\mathrm{min}\{\xij{01}^2, \xij{02}^2, \xij{21}^2 \})$. 

The strong coupling in the coordinate space is evaluated as 
\begin{equation}
\label{eq:running_coupling}
    \as(\xij{ij}^2) = \frac{4 \pi}{ \beta_0 \ln \left[ \left( \frac{\mu_0^2}{\lqcd^2}\right)^{1/c} + \left(\frac{4 C^2}{\xij{ij}^2 \lqcd^2}\right)^{1/c} \right]^c }.
\end{equation}
Here $\Lambda_\mathrm{QCD} = 0.241 \, \mathrm{GeV}$, $\beta = (11 \nc - 2N_f)/3$, and we use $N_f=3$ in this work. Based on~\cite{Kovchegov:2006vj,Lappi:2012vw}, the expected value for $\csq$ is  $C^2=e^{-2\gamma_{E}}\approx0.32$. This work, however, takes $C^2$ as a fit parameter to describe the uncertainty related to the running coupling prescriptions and to account for higher order corrections. Note that although we also include the charm quark contribution, we use $N_f=3$ at all transverse distance scales for simplicity (as in Ref.~\cite{Hanninen:2022gje}) as it is not obvious how $N_f$ should depend on this coordinate space scale\footnote{One potential prescription has been suggested in Ref.~\cite{Albacete:2010sy}}.

The KCBK equation~\eqref{eq:kcbk} is non-local in rapidity, with the rapidity shift
\begin{equation}
    \label{eq:kcbk-shift}
    \Delta_{012} = \max \left\lbrace 0, \ln \frac{\min\{ \xt_{02}^2,\xt_{21}^2 \} }{\xt_{01}^2} \right\rbrace.
\end{equation}
Following again Refs.~\cite{Beuf:2020dxl,Hanninen:2022gje}, we initialize the BK evolution at the kinematical limit $Y=0$ (corresponding to the ``$Y_{0,\mathrm{BK}}=0$'' setup of~\cite{Beuf:2020dxl}), and solve the equation numerically using the Euler method.

The initial dipole is parametrized using a McLerran-Venugopalan (MV) model~\cite{McLerran:1993ni} inspired parametrization, similarly as in Refs.~\cite{Lappi:2013zma,Albacete:2010sy,Beuf:2020dxl}:
\begin{multline}
    \label{eq:bk-ic}
    S_{01}(Y=0)  =
    \exp 
        \left[ 
            -\frac{\left(\xij{ij}^2Q_{s,0}^2\right)^\gamma}{4} \right.
             \\
            \left. \times \ln \left( \frac{1}{|\xij{ij}| \Lambda_\text{QCD}}
             +  e_c \cdot e   \right)   \right]. 
\end{multline}
Here the free parameters are the anomalous dimension, $\gamma$, infrared regulator, $e_c$, and $\qso^2$ controlling the initial saturation scale.
 
In principle it would be preferable to take $\gamma=1$ as in the MV model, as that forces the momentum space dipole amplitude (unintegrated gluon distribution) to be positive definite~\cite{Giraud:2016lgg}, and enables a more natural generalization for the dipole-nucleus scattering in the optical Glauber model~\cite{Lappi:2013zma}.
However, with such a parametrization for the initial condition we did not find any good fit. As such, we use the so-called MV$^\gamma$ parametrization from now on, fixing $e_c=1$ and considering $\gamma$ as a free parameter. 

The dipole amplitudes $N_{01}$ and $N_{012}$ in convolutions~\eqref{eq:sigma_qqg} and~\eqref{eq:sigma_qq}   satisfy the BK equation. Following the previous NLO fit of Ref.~\cite{Beuf:2020dxl}, when calculating the dipole term $\sigmaltdip$ the dipole amplitude is evaluated at $Y=\ln (1/x)$. In the case of $q\bar qg$-target interaction in $\sigmaltqgu$, stability of the factorization scheme requires the evolution rapidity to depend also on the gluon momentum fraction $z_2$~\cite{Beuf:2014uia,Ducloue:2016shw,Iancu:2016vyg,Ducloue:2017mpb,Ducloue:2017ftk}, and it becomes
\begin{equation}
\label{eq:y_z2}
Y \equiv \ln \frac{k^+}{P^+} = \ln \frac{W^2 z_2}{Q_0^2}.
\end{equation}
Here $k^+ = z_2 q^+$ is the gluon plus momentum, $q^+$ the large photon plus momentum  and $P^+ = Q_0^2/(2P^-)$.
Note that as already discussed above, when the $\sigmaltdip$ term is derived, an integral over $z_2$ from $z_2=0$ to $z_2=1$ is already performed analytically, and as such it is not possible to use the same $z_2$ dependent evolution rapidity in that term\footnote{The $z_2$ dependence to the $\sigmaltdip$
 has been restored in the massless limit in Ref.~\cite{Hanninen:2021byo}.}.

\section{Bayesian inference setup}
\label{sec:bayesian}

The inverse problem is a common challenge in high-energy particle physics phenomenology, where the underlying cause is extracted from observed effects, instead of the usual forward mindset of using theory to predict observations. Given the non-perturbative nature of the initial dipole-proton amplitude, it can only be constrained using available experimental data. 
The experimental datasets used in this work are the combined reduced total cross section~\cite{H1:2015ubc} and the reduced charm production cross section~\cite{H1:2018flt} measured in deep inelastic $ep$ scattering at HERA. In this analysis, we include data from the kinematical domain $2.0 \ \gev^2 \leq Q^2 < 50.0 \ \gev^2$ and $x \leq 0.01$ which gives us a total of 404 data points for the total reduced cross section and 34 data points for the charm contribution. The lower limit in $Q^2$ ensures the validity of the perturbative calculation, and the upper limit excludes the part of the phase space where Dokshitzer-Gribov-Lipatov-Altarelli-Parisi (DGLAP) dynamics would dominate.

Bayesian inference is a statistical approach that addresses the inverse problem in a probabilistic manner. The inference setup used in this work is equivalent to what we used in Ref.~\cite{Casuga:2023dcf} and is summarized here for completeness.

The posterior function  $P(\ymodel | \yexp)$ is the probability that a model $\ymodel$, at parameter point $\btheta$, is true in light of experimental data $\yexp$. Bayes' theorem inverts the likelihood function, $P(\yexp | \ymodel)$, to find the posterior
\begin{equation}
    P(\ymodel | \yexp) \propto P(\yexp | \ymodel) P(\btheta).
    \label{eq:bayesian}
\end{equation}
The likelihood function, $P(\yexp | \ymodel)$, is the degree that the model describes the observed data at a given parameter point. The log likelihood function can be written as
\begin{multline}
    \label{eq:loglikelihood}
    \log P(\yexp| \ymodel)) 
    \\ 
    = -\frac{1}{2} \left[ \deltay^{T} \Sigma^{-1}(\btheta) \deltay + \log ( 2\pi \ \mathrm{det}(\Sigma)) \right],
\end{multline}
where $\deltay = \ymodel - \yexp $, while $\Sigma = \Sigma_\mathrm{exp} + \Sigma_\mathrm{model}(\btheta)$. This work accounts for both uncorrelated and correlated systematic uncertainties present in the experimental data, where  $\Sigma_\mathrm{exp}$ is the full experimental covariance  matrix. The theoretical covariance  matrix provided by the Gaussian process emulator described below is represented by $\Sigma_\mathrm{model}$. 

The prior $P(\btheta)$ encodes knowledge, like bounds of the parameter space, about the model without the bias introduced by the experiment. In this work we consider a uniform prior, i.e. $P(\btheta)=1$ in the allowed region of the parameter space and $0$ outside.

Sampling algorithms belonging to the Markov Chain Monte Carlo (MCMC) class estimate the posterior and find the target distribution once the Markov Chain has reached equilibrium. An example of such is the Metropolis-Hastings (MH) algorithm. Starting from an initial position $\btheta_0$, a random point in the vicinity of the original point within the parameter space is selected as the proposed next step $\btheta_1$. The probability that the proposal is accepted is based on the ratio $P(\mathbf{y}(\btheta_{i+1})| \yexp)/P(\mathbf{y}(\btheta_{i})| \yexp)$. Eventually, the chain will converge to areas of high posterior and samples from this chain will make up the target posterior distribution. 

The MCMC algorithm, in this study, evaluates the posterior at around $\sim 10^6$ points or more in the parameter space. Hence, a Gaussian process emulator (GPE) is used in place of the model (the NLO cross section calculation) as a more efficient substitute. An estimate from the GPE is a function drawn from a Gaussian process that is characterized by a covariance kernel. We use a combination of the radial basis function and the Matérn kernels to form a more flexible covariance kernel. The hyperparameters describing the kernel are then optimized by training the emulator. The training data comprises of model calculations for parameter vectors chosen through a latin hypercube algorithm, ensuring an even spread throughout the parameter space. Principal component analysis further reduces the dimensionality of the training data and enables independent training of a GPE for each principal component. 
The  covariance matrix of the GPE $\Sigma_{\mathrm{emulator}}$, which describes the GPE's estimated uncertainty,  will represent the theoretical uncertainty of the model, $\Sigma_{\mathrm{model}} = \Sigma_{\mathrm{emulator}}$.

This work employs the $\texttt{emcee}$ package \cite{emcee} that implements an affine-invariant ensemble sampler of MH walkers, whose positions are updated based on the posterior evaluated by the other walkers. The GPEs are  implemented using the $\texttt{sklearn}$ package~\cite{scikit-learn}. We train separate emulators for the reduced cross section and for the charm production using training data at 500 parameter points, and 6 principal components that capture more than $99\%$ of the model variance. The GPE and MCMC sampler work hand in hand to explore the parameter space and extract the target posterior distribution. 

\begin{figure}[tb]
    \centering
    \includegraphics[width=0.49\linewidth]{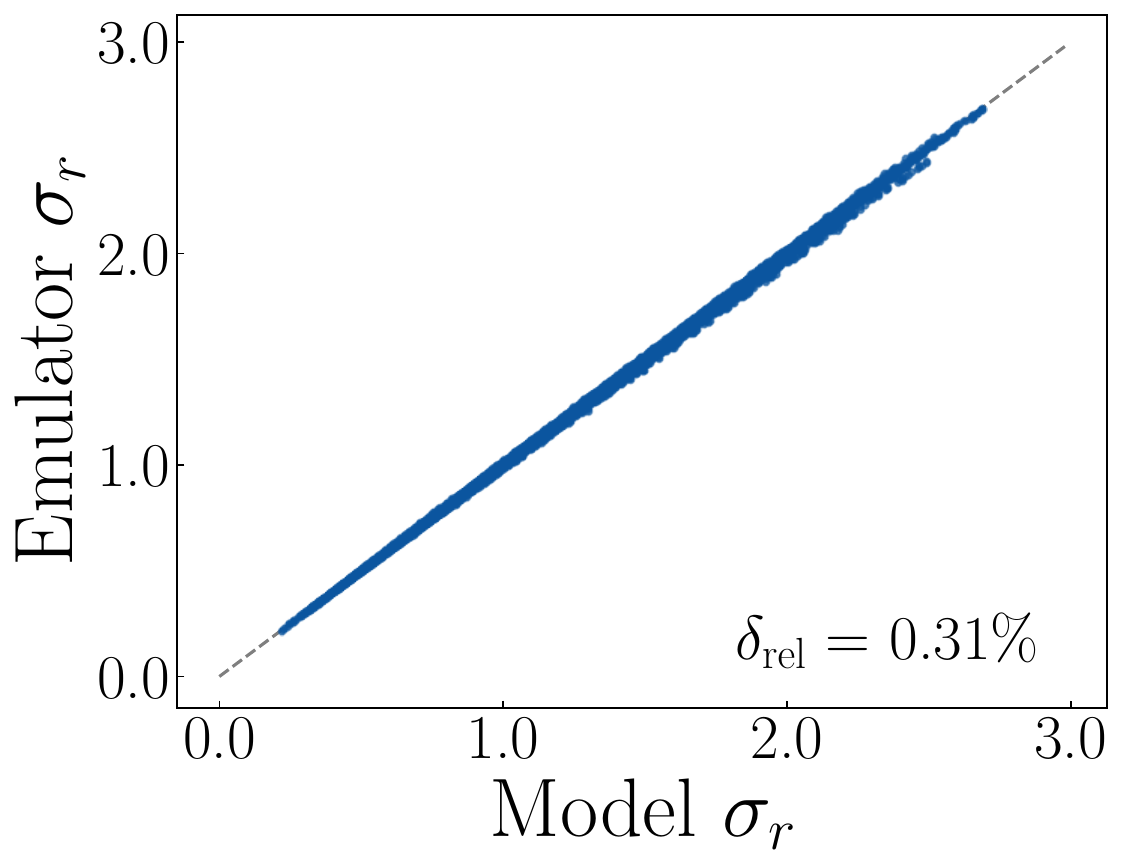}
    \includegraphics[width=0.49\linewidth]{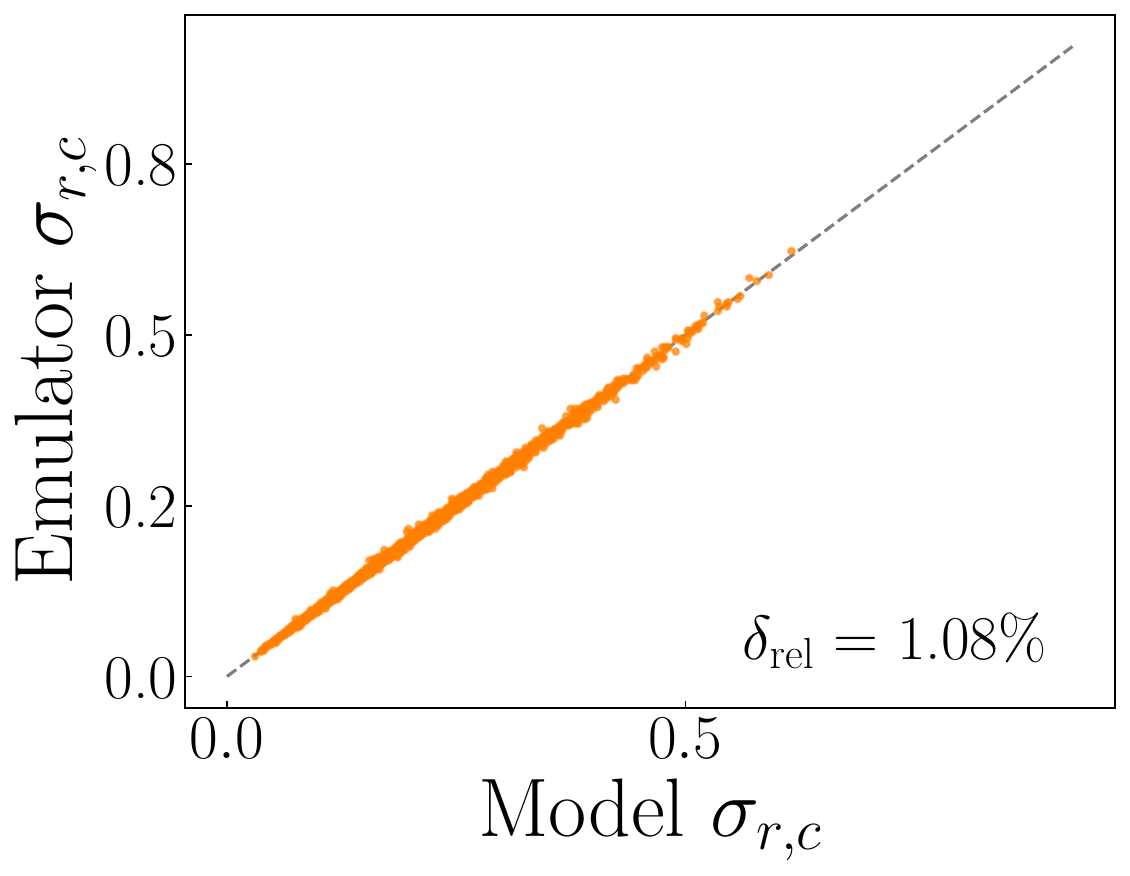}
    \caption{Comparison between the emulator prediction and the model calculation for the setup using the Balitsky+smallest dipole running coupling prescription.}
    \label{fig:validation_balsd}
\end{figure}

\begin{figure}[tb]
    \centering
    \includegraphics[width=0.49\linewidth]{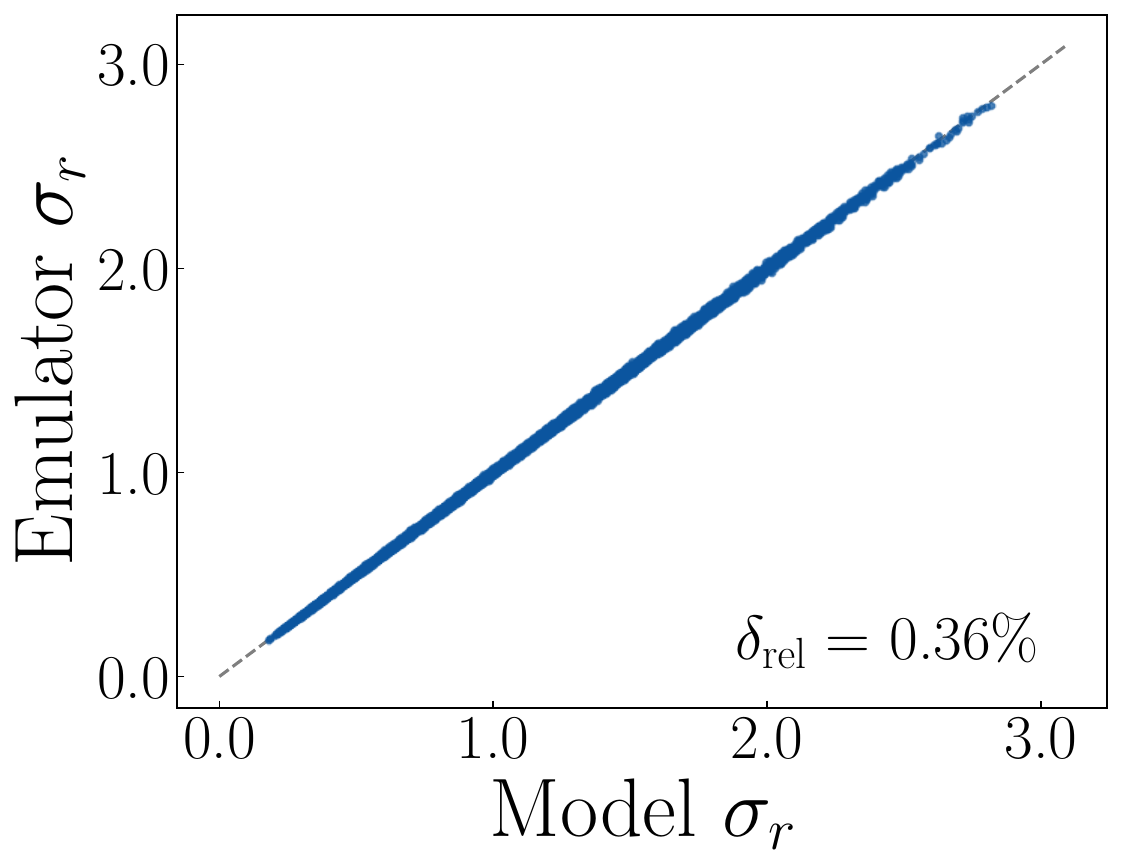}
    \includegraphics[width=0.49\linewidth]{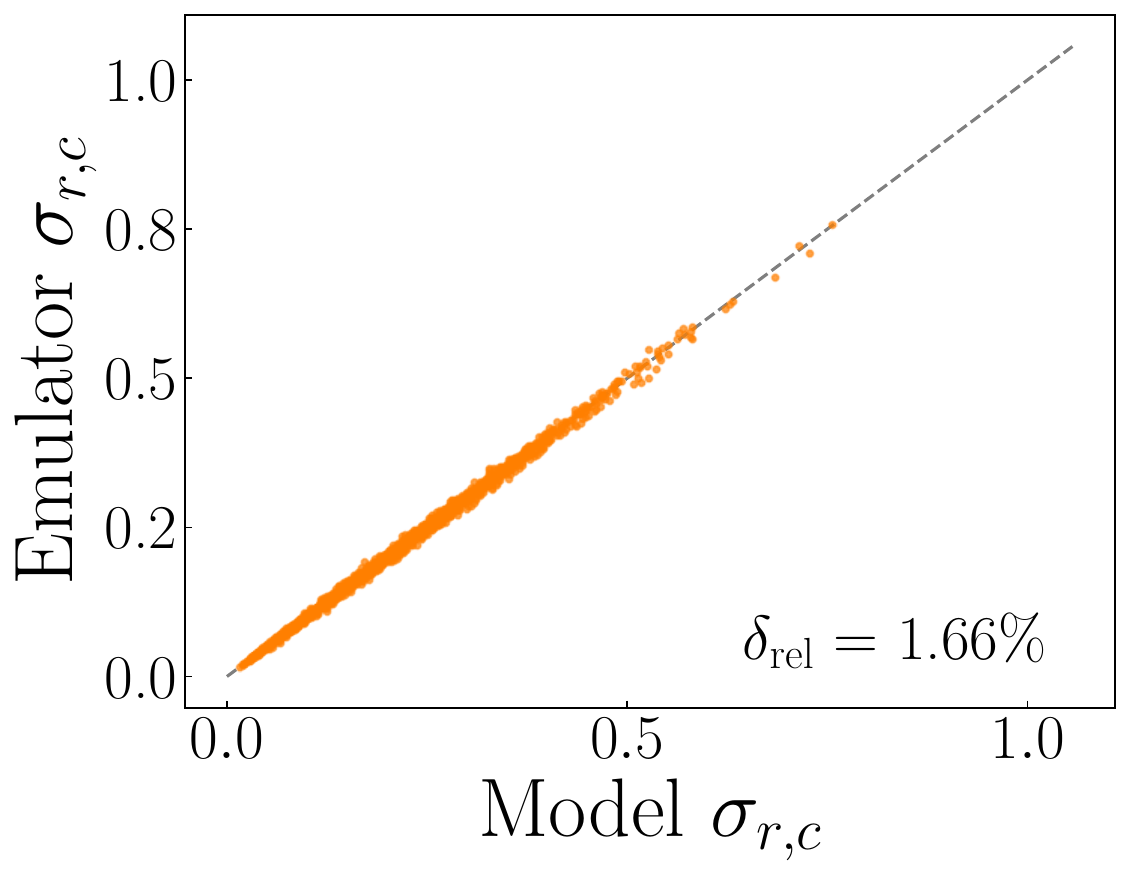}
    \caption{Comparison between the emulator prediction and the model calculation for the setup using the parent dipole running coupling.}
    \label{fig:validation_pd}
\end{figure}

\begin{figure}[tb]
    \centering
    \subfloat[Balitsky+smallest dipole]{
    \includegraphics[width=0.49\linewidth]{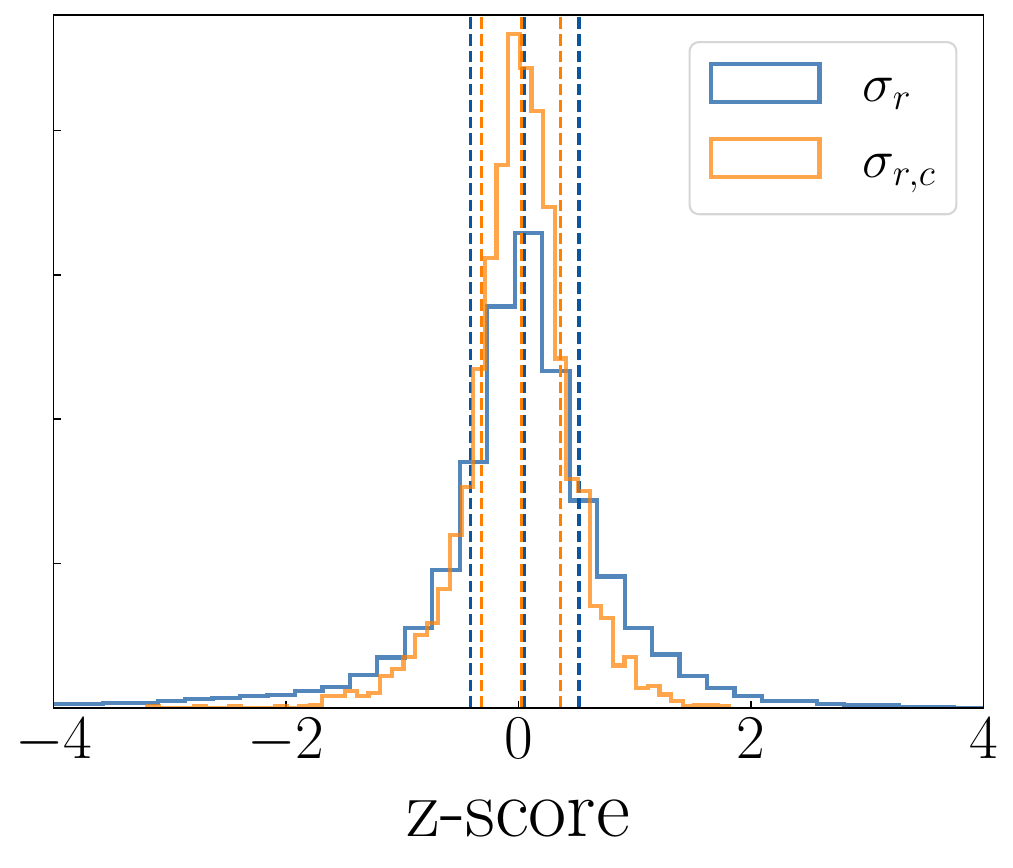}
    }
    \subfloat[Parent dipole]{
    \includegraphics[width=0.49\linewidth]{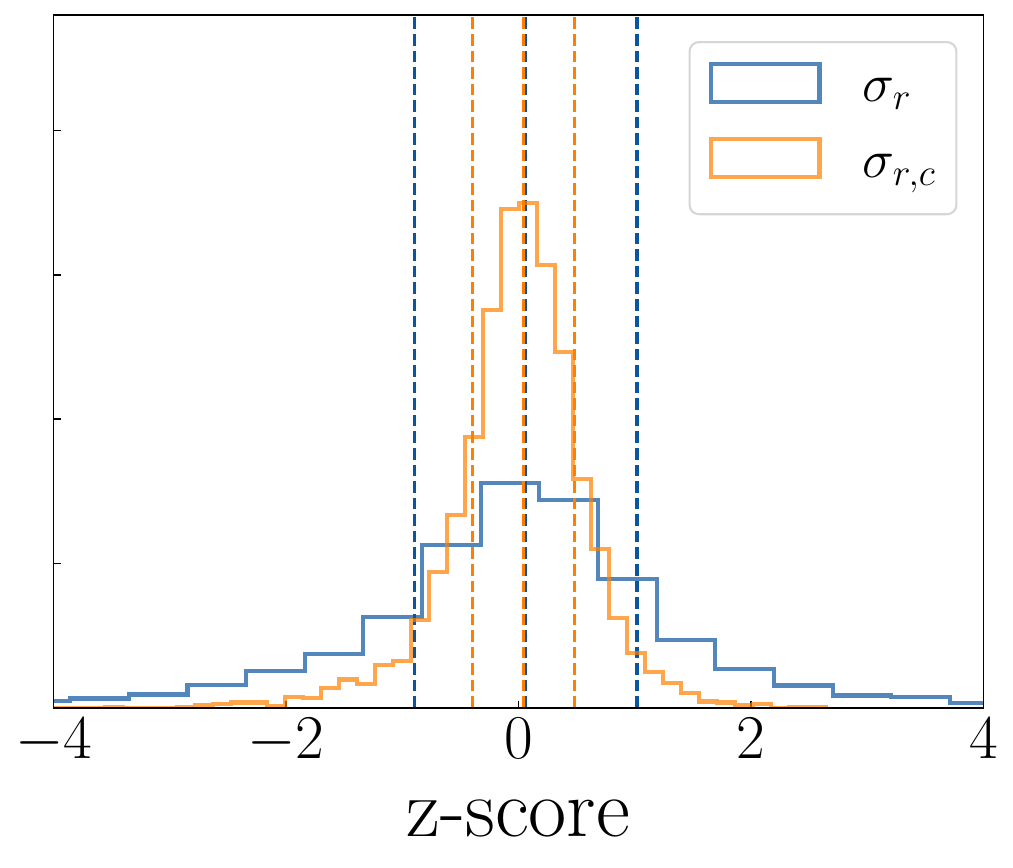}
    }
    \caption{The $z$-score distribution over 100 validation points sampled from the prior. The vertical lines indicate the one-standard-deviation interval. }
    \label{fig:zscore}
\end{figure}

\begin{figure*}[tb]
    \centering
     \subfloat[Balitsky+smallest dipole coupling]  {%
        \includegraphics[width=\linewidth]{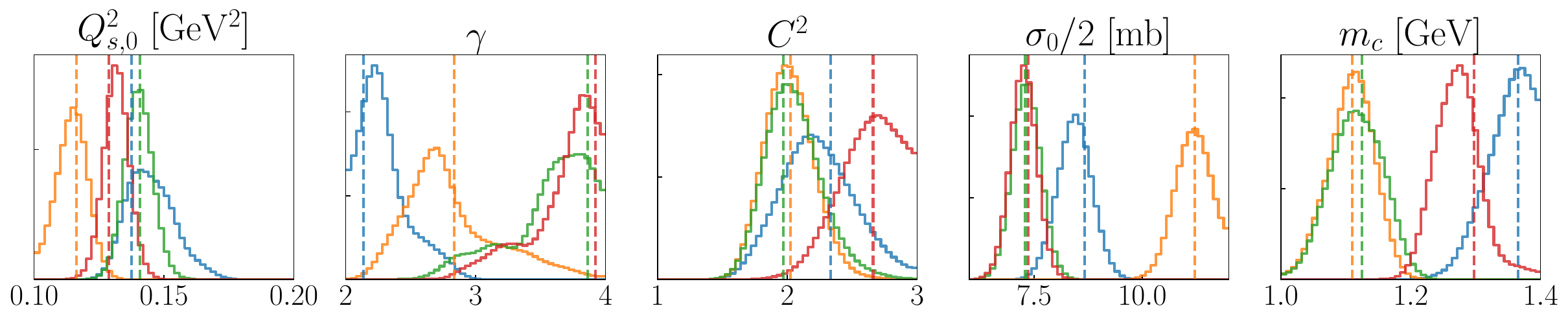}
    } \\
    \subfloat[Parent dipole coupling] 
    {%
        \includegraphics[width=\linewidth]{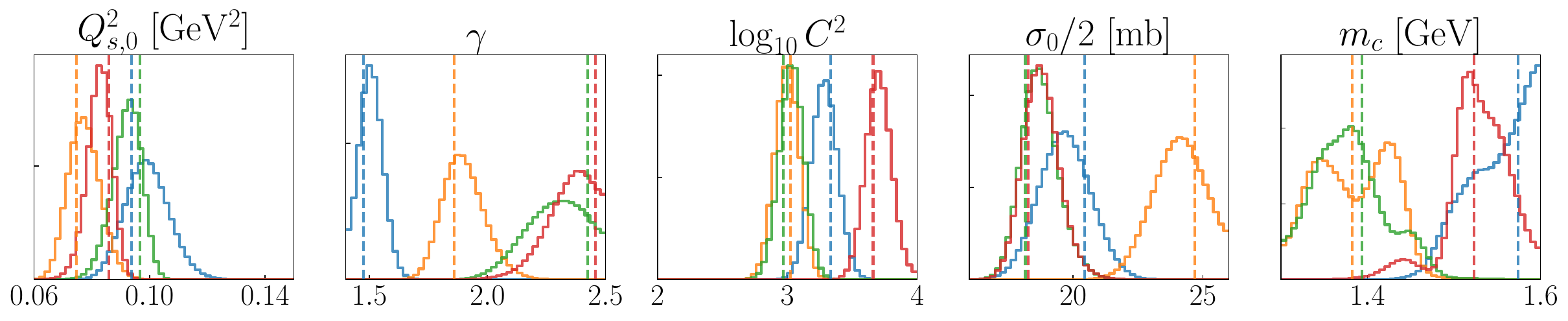}
    }
    \caption{Results from the closure test, comparison between extracted posterior distributions (solid lines) and corresponding true parameter values (dashed lines)}
    \label{fig:closure_test}
\end{figure*}

To validate the emulators, we quantify the emulator accuracies and determine if the uncertainty estimates are realistic. 
First, comparisons between the emulator and the model when calculating the total reduced cross section $\sigma_r$ and the charm production cross section $\sigma_{r,c}$ are shown for the Balitsky+smallest dipole and the parent dipole setups in Figs.~\ref{fig:validation_balsd} and~\ref{fig:validation_pd}, respectively. Overall we find an excellent agreement, and the average relative uncertainty $\delta_\mathrm{rel}$ shown in the figures in all cases is clearly smaller than the typical uncertainty in the HERA data. 

The uncertainty estimate provided by the emulator is validated by computing the $z$ score, which is a distribution of the difference between the emulator prediction and the actual model output, normalized by the emulator uncertainty estimate. The $z$-score distribution for both running coupling schemes is shown in Fig.~\ref{fig:zscore}.  
For a perfect emulator this distribution has a zero mean and unit width. In our case, we observe that our GPEs slightly overestimate the emulator uncertainty, as reflected by the standard deviations indicated by dashed vertical lines being typically less than 1. Overall we conclude that the emulators are very accurate, and the emulator uncertainties are also realistic and can be used in the MCMC sampling phase.

Finally we validate the MCMC sampler and the overall Bayesian inference framework by performing a closure test. This test constrains the model to an artificially generated data constructed from known input parameters, here considered as the \textit{true} values.
Each datapoint is then assigned an uncertainty such that the relative uncertainty matches the HERA data.
The established GPE and MCMC sampler should then be able to recover a posterior distribution that is aligned with the true values. We do such a test for 4 distinct parameter points chosen, and subsequently removed, from the emulator training set. Fig. \ref{fig:closure_test} demonstrates a close alignment between the peaks of the posterior distributions and the known true values (vertical dashed lines). Such agreement allows us to be confident on the robustness and reliability of our own Bayesian setup in extracting target model parameters.

% ---- % ---- % ---- % ---- % ---- % ---- % 

\section{Results}
\label{sec:results}

\begin{figure*}[ht]
    \centering
    \includegraphics[width=\linewidth]{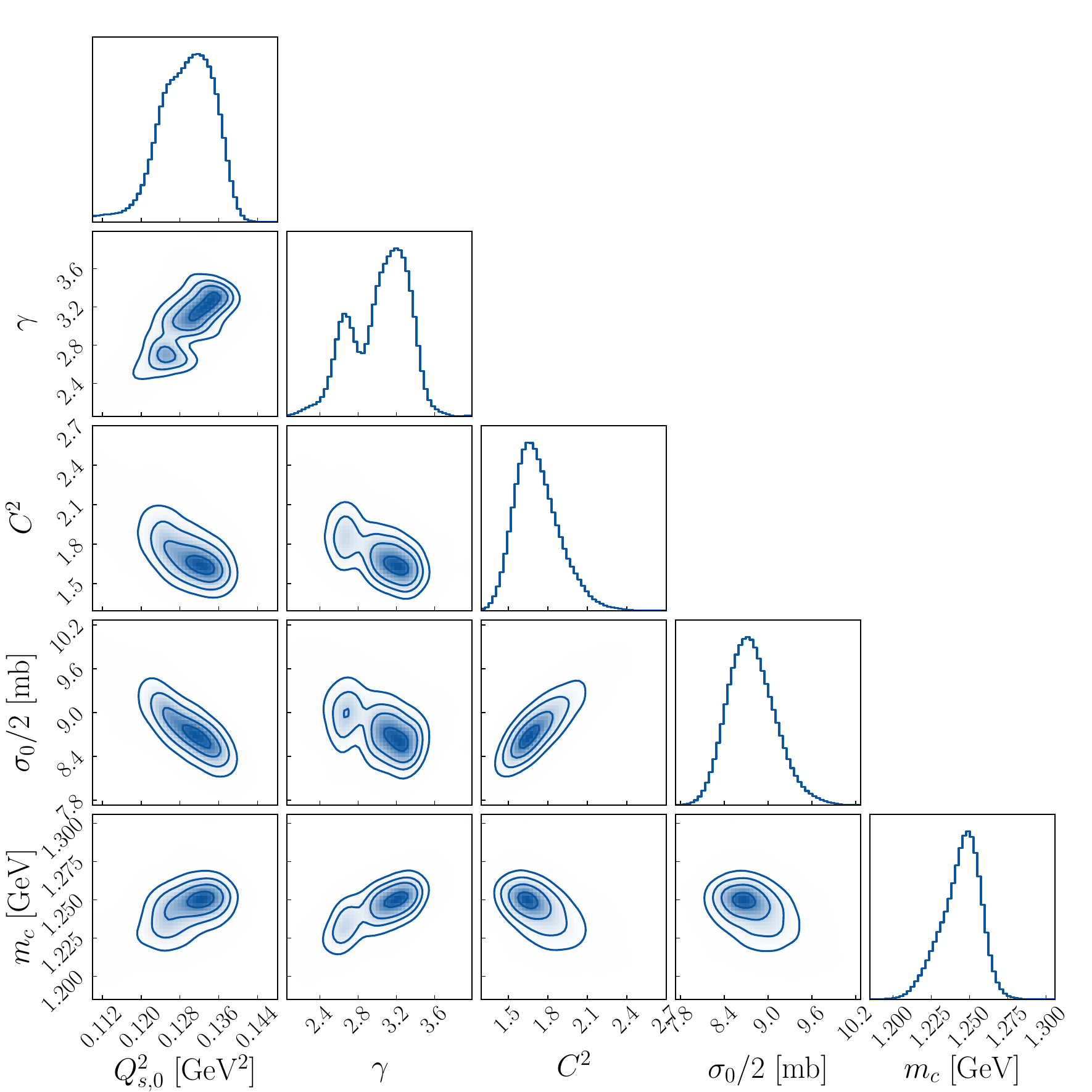}
    \caption{1- and 2-dimensional projections of the posterior probability distribution for the Balitsky+smallest dipole running coupling setup. } 
    \label{fig:kcbk_balsd_corner}
\end{figure*}

\begin{figure*}[ht]
    \centering
    \includegraphics[width=\linewidth]{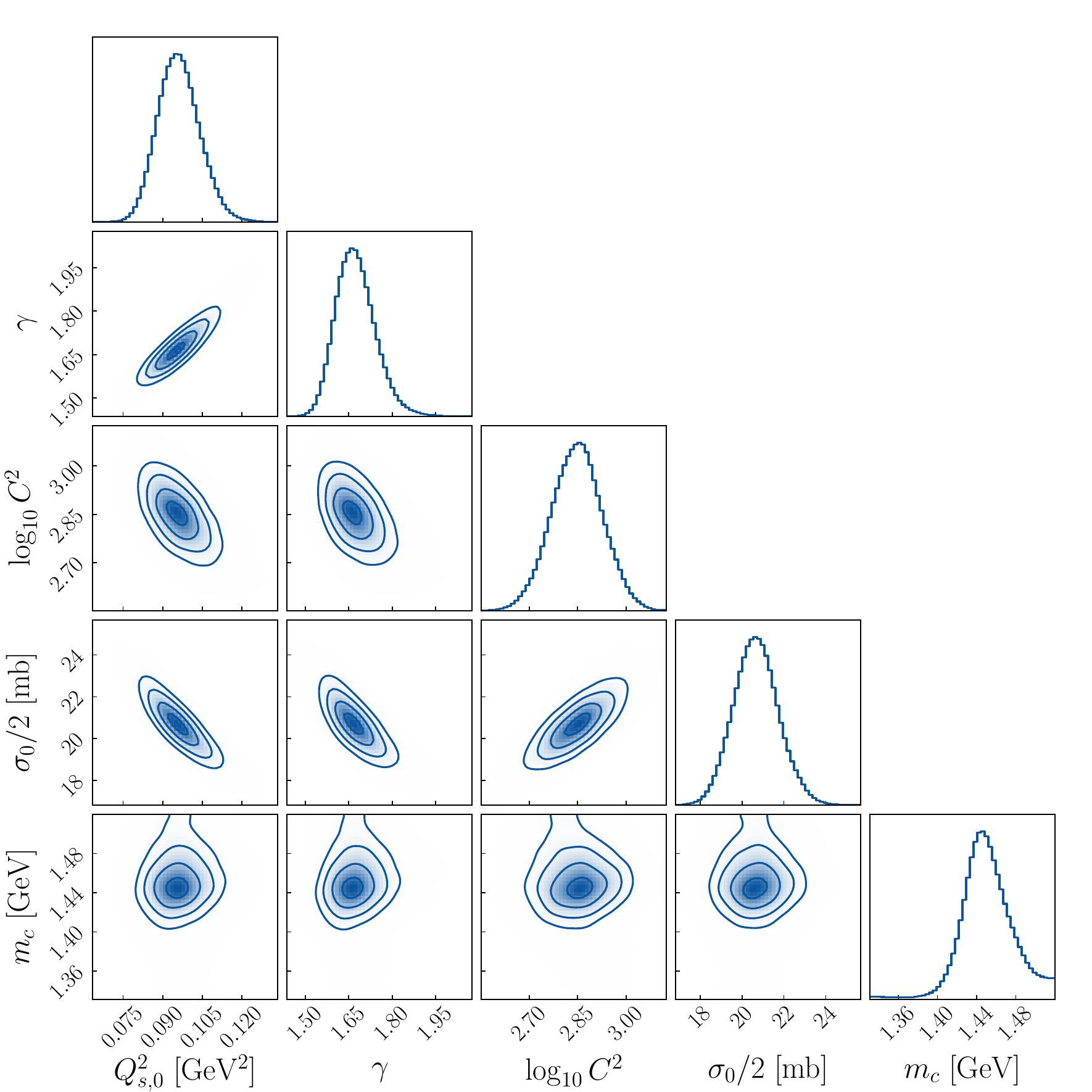}
    \caption{1- and 2-dimensional projections of the posterior probability distribution for the  parent dipole running coupling setup}
    \label{fig:kcbk_pd_corner}
\end{figure*}

The model parameters $\qso^{2}$, $\g$, $\csq$, $\initsig$, and $m_c$  are simultaneously constrained through Bayesian inference using HERA combined data for the total DIS cross section and for the charm production. The charm mass, $\mc$, in the pole mass scheme used in the calculation of Ref.~\cite{Beuf:2021srj}, is also considered to be a fit parameter as in Ref.~\cite{Hanninen:2022gje}. 

\renewcommand{\arraystretch}{1.5} 
\begin{table*}[ht]
    \centering
    \begin{tabular}{|p{7.0cm}|wc{2.2cm}|wc{2.2cm}|wc{2.2cm}|wc{2.2cm}|}
    \hline
        \multirow{2}{*}{\textbf{Parameter Description}}  & \multicolumn{2}{c|}{\textbf{KCBK + Bal+SD}} & \multicolumn{2}{c|}{\textbf{KCBK + parent dipole}}  \\ \cline{2-5}
          & Median $\pm \ 2\sigma$  & MAP & Median $\pm \ 2\sigma$  & MAP \\
    \hline
    \hline
        Initial scale, $\qso^2 \ \mathrm{[GeV}^{2}]$ &  $0.130_{-0.012}^{+0.008}$ & 0.124 &  $0.096_{-0.014}^{+0.016}$ & 0.090 \\
        
        Anomalous dimension, $\gamma$ & $3.08_{-0.64}^{+0.38}$ & 3.23 & $1.67_{-0.11}^{+0.15}$ & 1.60 \\

        Running coupling scale, $C^{2}$  & $1.70_{-0.25}^{+0.41}$ & 1.74 & $705_{-209}^{+289}$ & 663 \\

        Proton transverse area, $\sigma_{0}/2 \ \mathrm{[mb]}$  & $8.75_{-0.53}^{+0.72}$ & 9.08 & $20.7_{-2.0}^{+2.1}$ & 20.7 \\

        Charm mass, $m_c \ \mathrm{[GeV]}$ & $1.25_{-0.03}^{+0.02}$ & 1.24 & $1.45_{-0.05}^{+0.06}$ & 1.40 \\

    \hline
    Saturation scale, $Q_s^{2}$, at $Y=\ln(1/0.01)$ $[\mathrm{GeV}^2]$ & 0.203 & 0.196 & 0.208 & 0.199 \\ 

    \hline
    \hline
    
    \multicolumn{5}{|l|}{\textbf{$\chi^2$/d.o.f. values}} \\
    \hline
        
    $\sigma_r$ data only & 1.18 & 1.18 & 1.12 & 1.09 \\ 

    $\sigma_{r,c}$ data only & 3.96 & 4.37 & 2.84 & 3.12 \\

    All data & 1.35 & 1.38 & 1.22 & 1.21 \\
    \hline 
    \hline
    \multicolumn{5}{|l|}{\textbf{Average $\chi^2$/d.o.f. over 100 samples}} \\
    \hline
    $\sigma_{r}$ data only & \multicolumn{2}{c|}{1.19} & \multicolumn{2}{c|}{1.17} \\
    $\sigma_{r,c}$  data only & \multicolumn{2}{c|}{3.80} & \multicolumn{2}{c|}{2.68} \\
        All data & \multicolumn{2}{c|}{1.38}  & \multicolumn{2}{c|}{1.25} \\
    \hline
    \end{tabular}
    \caption{MAP and median values for the BK initial condition parametrization obtained using the Balitsky+smallest dipole (labelled as Bal+SD) and parent dipole running coupling schemes. 
    The uncertainty estimates within the 95\% credible intervals are also shown, as well as $\chi^2/\mathrm{d.o.f.}$ values when compared different datasets and when including all HERA data, and saturation scales at $Y=\ln \frac{1}{0.01}$ defined as $N(r^2=2/Q_s^2)=1-e^{-1/2}$. }
    \label{tab:medianmap}
\end{table*}

The posterior distribution of the model parameters obtained in the setup using the Balitsky+smallest dipole running coupling prescription is shown in Fig.~\ref{fig:kcbk_balsd_corner}, and similar results obtained using the parent dipole coupling are presented in Fig. \ref{fig:kcbk_pd_corner}.
Here the diagonal panels show the marginal distribution, and off-diagonal panels illustrate correlations between the two model parameters.
In both setups all model parameters are tightly constrained by the HERA data. 

The maximum-a-posteriori (MAP) parametrizations resulting in the highest likelihoods are shown in Table~\ref{tab:medianmap}, along with the $\chi^2/\mathrm{d.o.f.}$ values of $\mathcal{O}(1)$ demonstrating that the description of the precise HERA data obtained with only five parameters is overall excellent. We also report the posterior median with uncertainty values covering the 95\% credible interval. Reported also are the corresponding values for the model-independent proton saturation scale, $Q_s^2$, at $Y=\ln(1/0.01)$ obtained by solving $N(r^2 = 2/Q_s^2) = 1 - e^{-1/2}$. Samples from the posterior distribution are available in Ref.~\cite{samples_zenodo} These model parameter vectors can be used to define the initial condition for the BK evolution which is a necessary input for all NLO calculations within the CGC framework, enabling one to propagate the initial condition uncertainty to the cross section level. For all the posterior samples, datafiles containing the corresponding evolved dipole amplitudes are provided as well.

Similarly as in Ref.~\cite{Hanninen:2022gje}, a better simultaneous description of both the total and charm production datasets is obtained if the parent dipole running coupling scheme is used. The total cross section is described approximatively equally well with both setups. When the theoretically better motivated Balitsky running coupling scheme~\cite{Balitsky:2006wa} is used, we have $\chi^2/\mathrm{d.o.f.}\sim 4$ when the best-fit values are compared with the charm production data. However, also in that case the agreement with the very precise world data can be considered to be good as illustrated by the combined $\chi^2/\mathrm{d.o.f.} \sim 1.4$ values (averaged over 100 samples). 

The different running coupling schemes result in somewhat different parametrizations. In particular, the proton transverse area $\sigma_0/2$  depends heavily on this scheme choice. The value obtained for $\sigma_0/2$ with the parent dipole prescription is more than twice the value obtained when the Balitsky+smallest dipole prescription is used. Similar systematics has been previously observed in Ref.~\cite{Beuf:2020dxl} when fitting the light quark contribution only. 

While both setups give similar results for $\gamma^*$-proton scattering, the differing proton transverse areas can significantly affect predictions for nuclear targets.
In the optical Glauber model, the nuclear saturation scale as a function of the impact parameter $\bt$ behaves as  ~$Q_{s,A}^2(\bt) \sim \frac{\sigma_0}{2} A T_A(\bt) Q_{s,0}^{2\gamma}$~\cite{Lappi:2013zma}. As such the magnitude of the non-linear effects in dipole-nucleus scattering depends rather heavily on the extracted proton size, see e.g. Ref.~\cite{Mantysaari:2023vfh} in the case of inclusive hadron production in proton-lead collisions. Here $A$ is the nuclear mass number and $T_A(\bt)$ is the transverse thickness profile normalized as $\int \dd[2]{\bt} T_A(\bt)=1$. As such, future nuclear DIS data, or other scattering processes measured e.g. at the LHC, can be expected to provide complementary constraints for the non-perturbative initial condition.

Complementary constraints for the proton size can also be obtained from exclusive processes that directly probe the target shape~\cite{Klein:2019qfb}, or from diffractive structure function measurements~\cite{Lappi:2023frf}. Depending on the form assumed for the proton geometry, exclusive $\mathrm{J}/\psi$ production data from HERA ~\cite{H1:2013okq,ZEUS:2002wfj} prefers $\sigma_0/2\sim 10\dots 20$ mb, with the lower value obtained with a Gaussian shape and the upper limit when a step function profile (which is not fully compatible with the measured spectra~\cite{Kowalski:2006hc}) is used. As such, we conclude (as in Ref.~\cite{Beuf:2020dxl}) that both setups are compatible with the vector meson production data, although those measurements  prefer the value obtained with the (more realistic) Balitsky+smallest dipole prescription.

The other main difference between the two running coupling schemes is on the value obtained for the running coupling scale, parametrized by $C^2$. As discussed in Sec.~\ref{subsec:evolution}, the Balitsky prescription generically results in slower evolution speed. For this reason, the fit with the parent dipole scheme features larger values of $\csq \sim 700$, as a larger $\csq$ slows  down the evolution, see Eq.~\eqref{eq:running_coupling}. When the Balitsky prescription is used, one obtains $C^2\sim 1.5$ (note that dependence on $C^2$ is logarithmic). It is still significantly larger than the $C^2=e^{-2\gamma_E}$ estimate discussed in Sec.~\ref{subsec:evolution}. However,  we also note that the full NLO BK equation is known to result in even slower evolution than the resummation-only approximation employed here~\cite{Lappi:2016fmu}, which should reduce the optimal value for $C^2$ further.

The charm mass is weakly correlated with other model parameters. This is expected, as it is most directly constrained by the charm production data. The values for the charm quark mass  $\mc$ obtained using both running coupling schemes are similar to what has been reported in Ref.~\cite{Hanninen:2022gje}.

The proton transverse area $\initsig$ (that acts as a normalization factor) is  correlated with $\csq$ and anticorrelated with $\qso^{2}$. These are the same correlations observed in the leading order posterior distribution in Ref.~\cite{Casuga:2023dcf}.  The anticorrelation between $\qso^2$ and $\initsig$ is because non-linear effects are weak in dipole-proton scattering, and as such the cross section scales as $\sim \frac{\sigma_0}{2} Q_{s,0}^{2\gamma}$. 

\begin{figure*}[ht]
    \subfloat[Total cross section]  {%
        \includegraphics[width=0.49\textwidth]{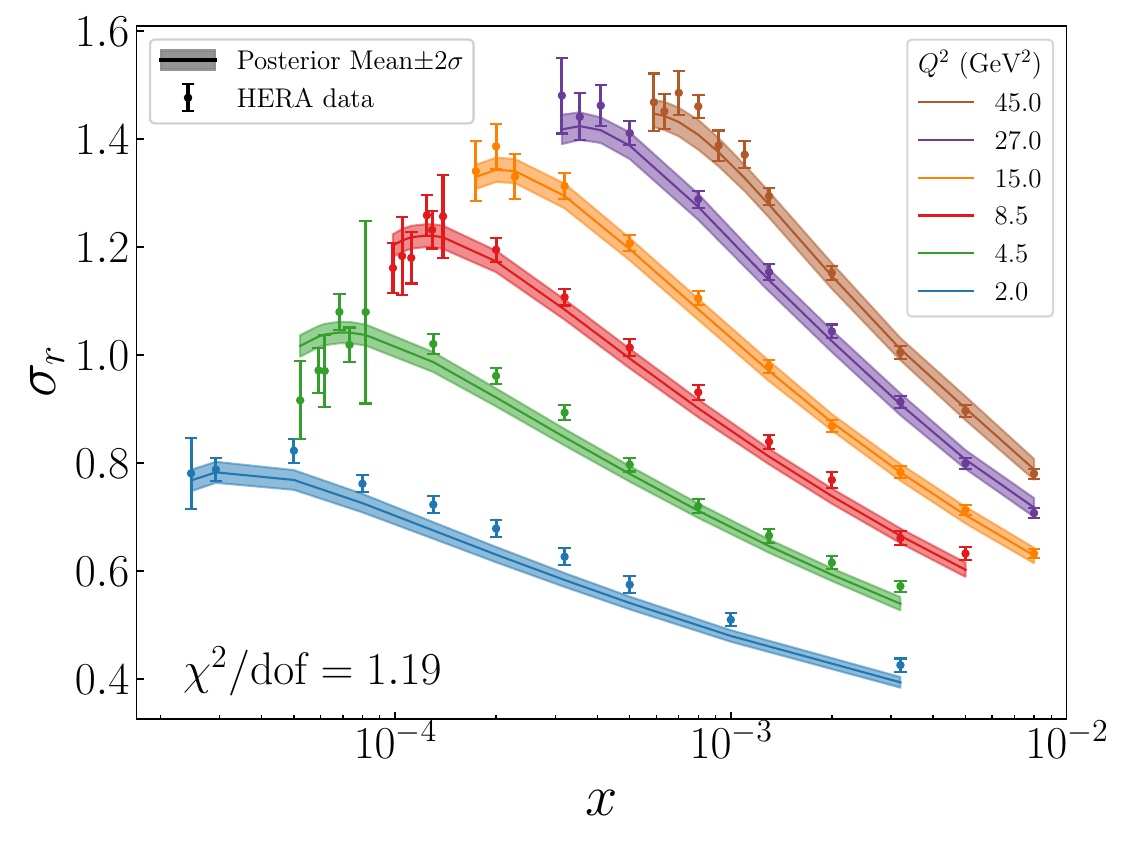}%
        \label{fig:}%
    }
    \hfill
    \subfloat[Charm production]{%
        \includegraphics[width=0.49\textwidth]{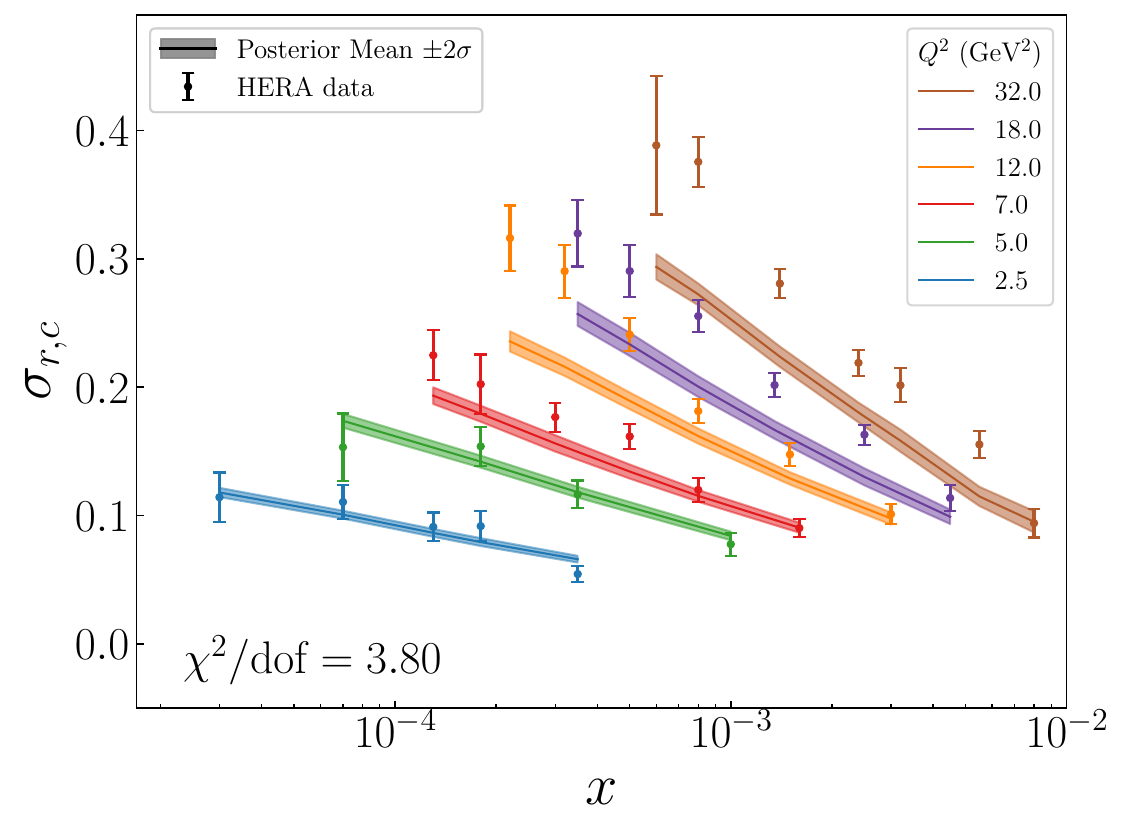}%
        \label{fig:}%
    }
    \caption{Reduced total cross section and charm production as a function of $x$, in selected $Q^2$ bins at $\sqrt{s} = 318 \ \gev$, calculated for posterior samples from fit using Balitsky+smallest dipole running coupling scheme compared to HERA data.
    } 
    \label{fig:balsd_data_comparison}
\end{figure*}

\begin{figure*}[ht]
    \subfloat[Total cross section]  {%
        \includegraphics[width=0.49\textwidth]{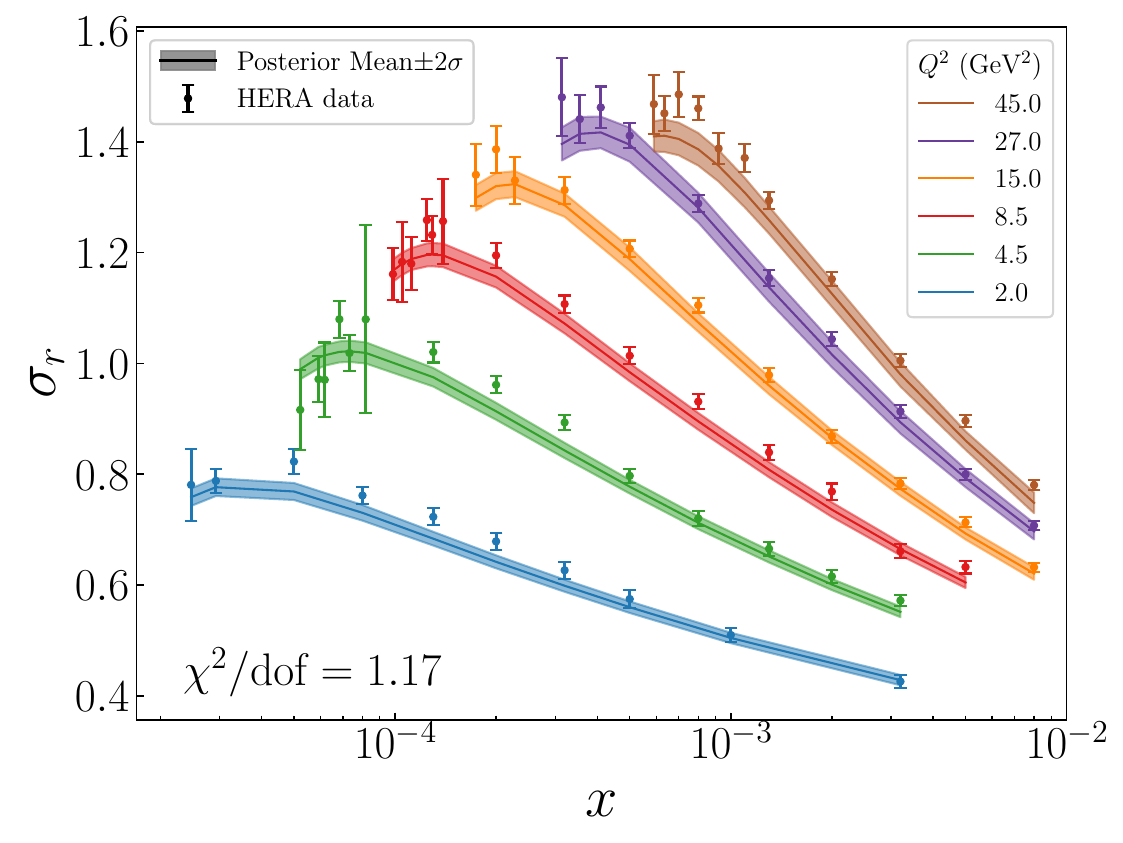}
    }
    \hfill
    \subfloat[Charm production]{%
        \includegraphics[width=0.49\textwidth]{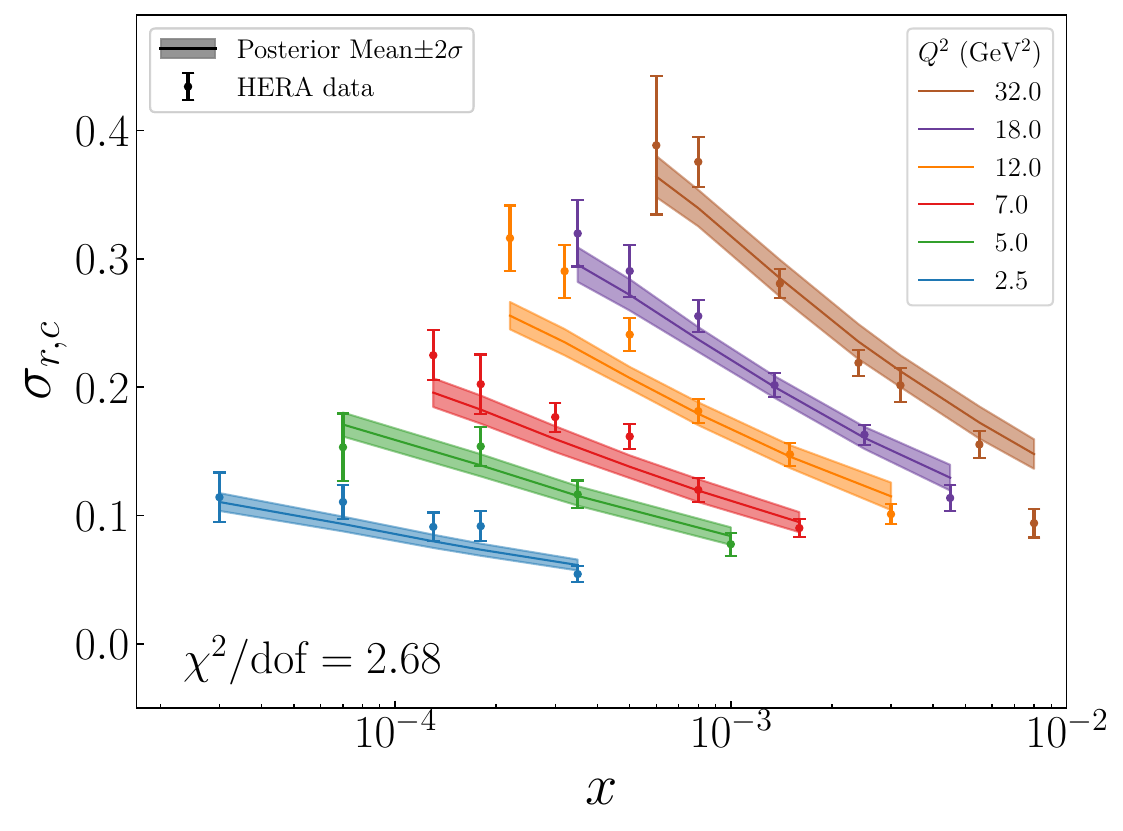}%
    }
     \caption{Reduced total cross section and charm production as a function of $x$, in selected $Q^2$ bins at $\sqrt{s} = 318 \ \gev$, calculated for posterior samples from fit using parent dipole running coupling scheme compared to HERA data.} 
    \label{fig:pd_data_comparison}
\end{figure*}

The quality of our fit is demonstrated in Figs.~\ref{fig:balsd_data_comparison} and~\ref{fig:pd_data_comparison} where the reduced cross section and charm production are calculated using samples from both posterior distributions, and compared to the HERA data. In addition to central (average) values, the $2\sigma$ uncertainty estimates are shown as well. As seen from the good $\chi^2/\mathrm{d.o.f.}$ values shown in the figure and quoted in Table~\ref{tab:medianmap}, as well, an excellent description of the total cross section data is obtained. 

The description of the charm production data is somewhat worse, with the cross section typically underestimated at small-$x$ and large $Q^2$. This tension between the two datasets is observed using both running coupling schemes, as the charm production data would prefer a faster $Q^2$ dependence. That would be obtained by having a smaller $\g$, which would then result in a worse description of the $Q^2$ dependence in the total cross section data. This tension is especially clear when the Balitsky+smallest dipole running coupling prescription is used. However, in both setups the global $\chi^2/\mathrm{d.o.f.}$ (calculated including all data) remains good, see Table~\ref{tab:medianmap}. We also note that inclusion of the correlated experimental uncertainties allows for some model deviation and the agreement with the data may not look as good as the $\chi^2/\mathrm{d.o.f.}$ value would suggest. 

\begin{figure*}
    \centering
    \subfloat[Balitsky+smallest dipole running coupling scheme]  {%
        \includegraphics[width=0.5\textwidth]{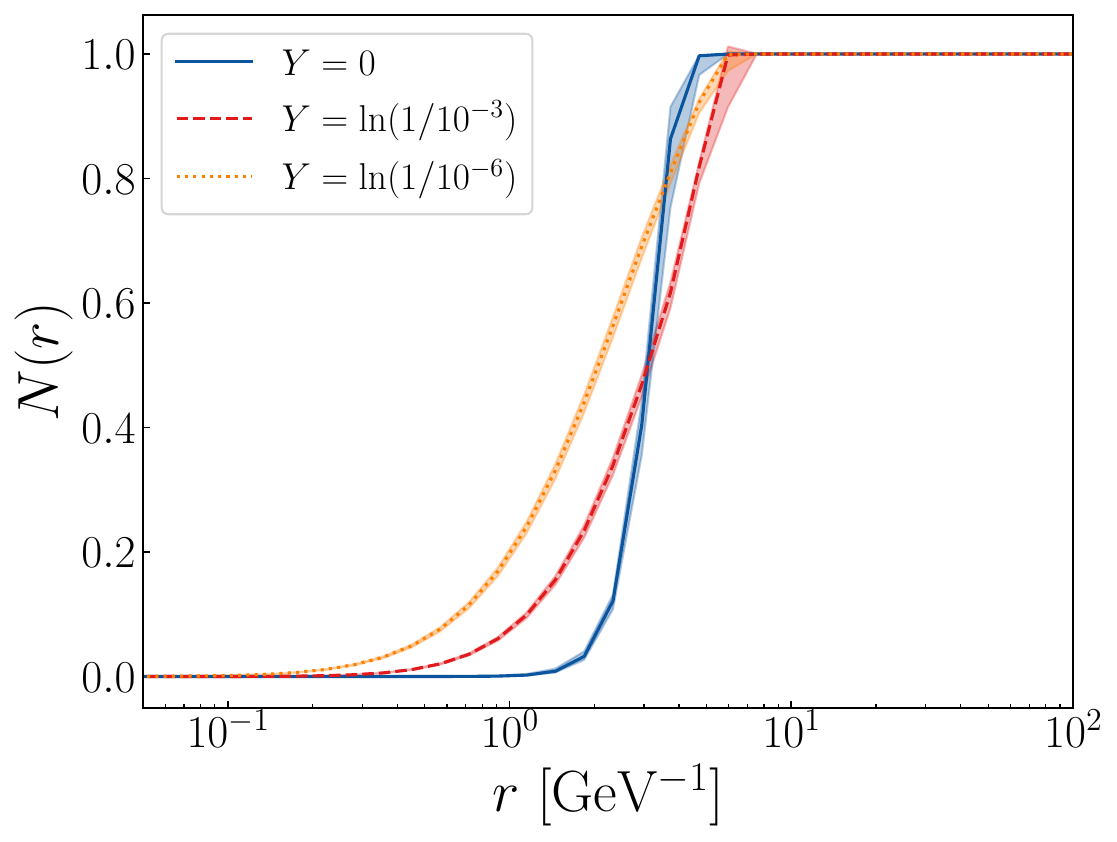}%
        \label{fig:dipole_balsd}%
    }
    \hfill
    \subfloat[Parent dipole running coupling scheme]{%
        \includegraphics[width=0.5\textwidth]{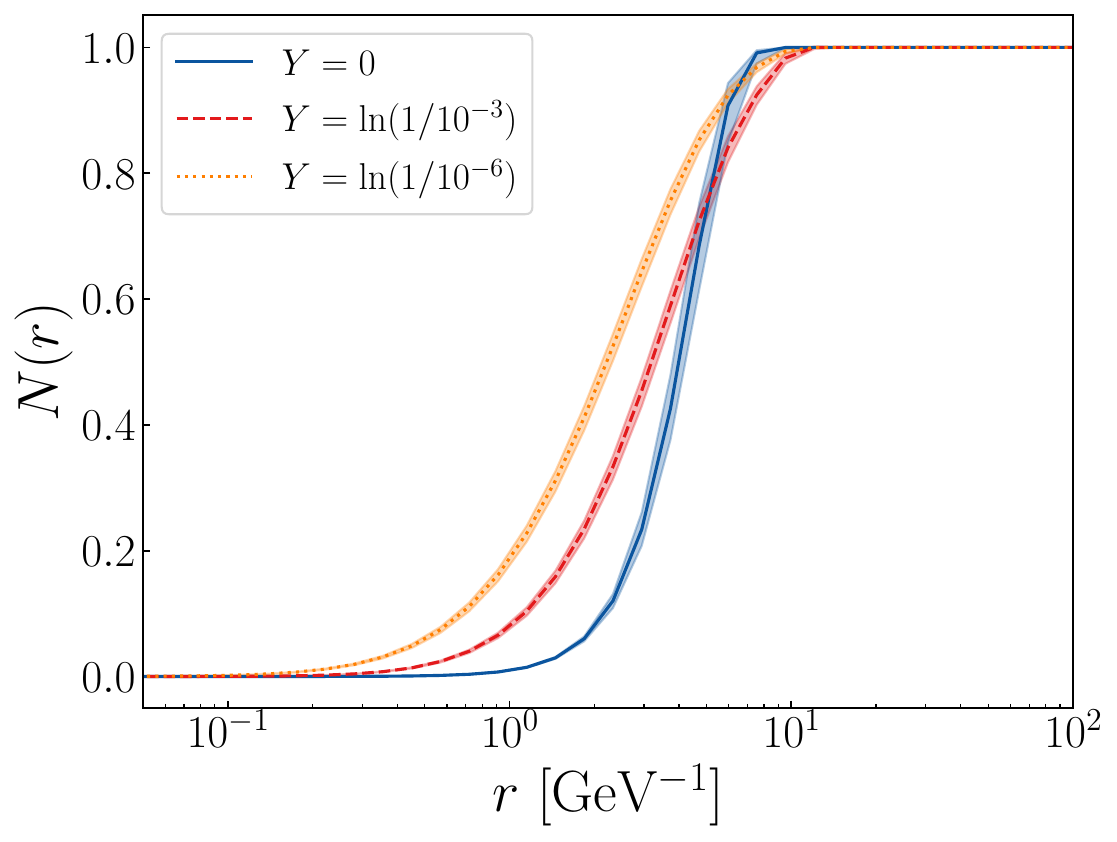}%
        \label{fig:dipole_pd}%
    }
    \caption{The dipole amplitude at initial condition, $Y=0$ and evolved towards $x=10^{-3}$ and $x=10^{-6}$ calculated for posterior samples from fit using Balitsky+smallest dipole or parent dipole schemes. The band describes the 2$\sigma$ uncertainty. } 
    \label{fig:dipole}
\end{figure*}

\begin{figure*}
    \subfloat[Balitsky+smallest dipole running coupling scheme]  {%
        \includegraphics[width=0.5\textwidth]{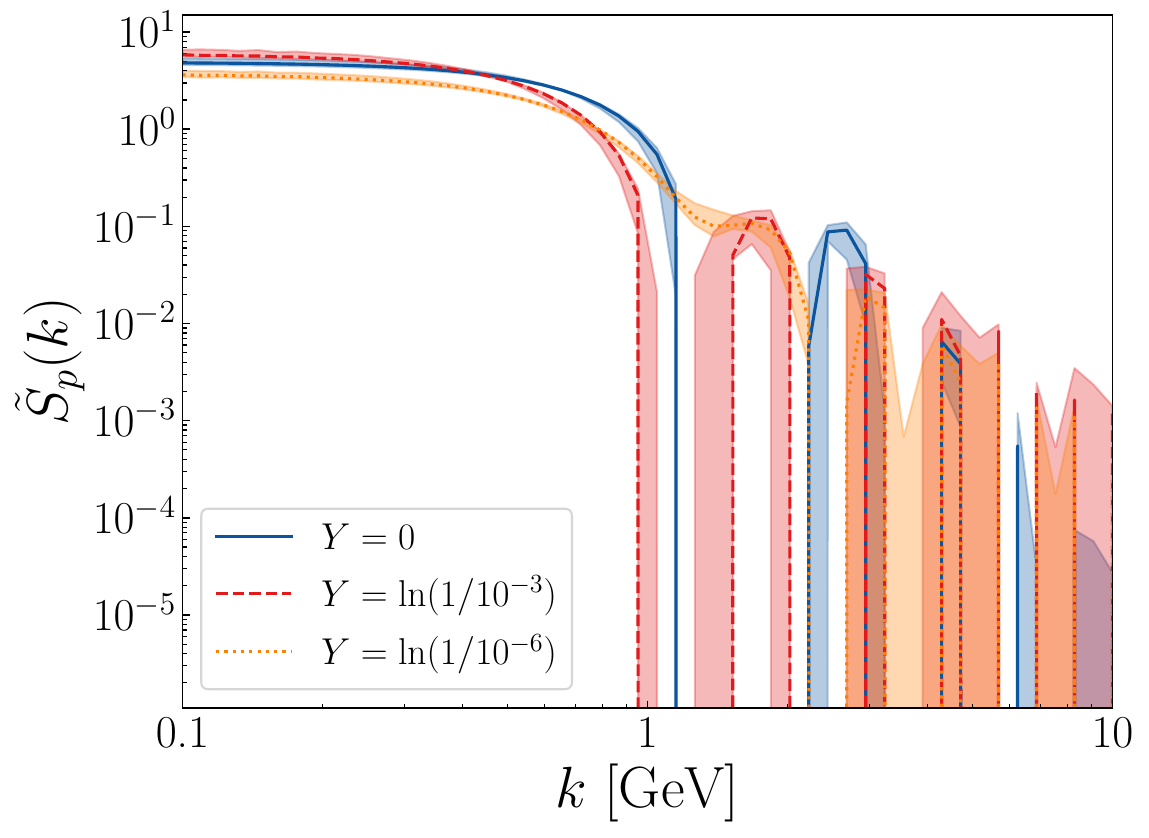}%
        \label{fig:2dft_kcbk_balsd}%
    }
    \hfill
    \subfloat[Parent dipole running coupling scheme]{%
        \includegraphics[width=0.5\textwidth]{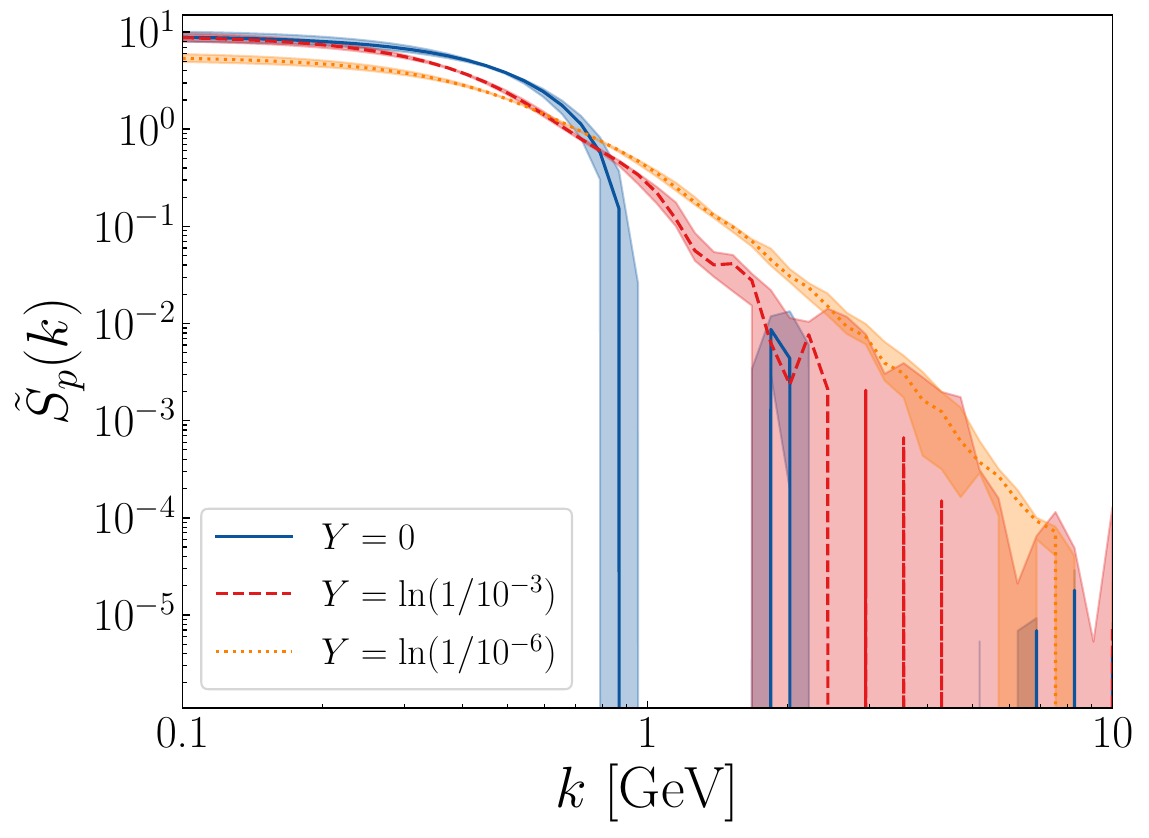}%
        \label{fig:2dft_kcbk_pd}%
    }
    \caption{2-dimensional fourier transform of the initial and evolved, to  $x=10^{-3}$ and $x=10^{-6}$, dipole calculated for posterior samples from fit using Balitsky+smallest dipole or parent dipole schemes. The band describes the $2\sigma$ uncertainty. }
    \label{fig:2dft}
\end{figure*}

The  dipole amplitudes at the initial condition and after $Y=\ln (1/10^{-3})$ and $Y=\ln(1/10^{-6})$ units of BK evolution are shown in Fig.~\ref{fig:dipole} with uncertainty estimates. The anomalous dimension $\gamma$ controls the slope of the dipole amplitude at small $\rt$ where $N \sim (\rt^2 \qso^2)^\gamma$, and due to the large $\gamma$ values obtained here (similarly as in previous analyses~\cite{Albacete:2009fh,Albacete:2010sy,Lappi:2013zma}),  the dipole amplitude is steeply falling at $r\lesssim 1/Q_s$. Although the resummed evolution does not significantly alter the shape of the dipole at small $r$~\cite{Lappi:2016fmu}, there is a clear transition towards a traveling-wave like solution at intermediate $r$. Especially with the Balitsky+smallest dipole prescription where the initial $\g$ is larger, this transition effect is still visible in the LHC kinematics corresponding to $Y=\ln (1/10^{-6})$.

In the MV-model based parametrizations, $\g>1$ results in negative unintegrated gluon distribution~\cite{Giraud:2016lgg}. This is demonstrated in Fig.~\ref{fig:2dft} for the 2-dimensional Fourier transform (2DFT) of the  dipole amplitude
\begin{equation}
    \tilde S_p(\kt) = \int \dd[2]{\rt} e^{i\kt \cdot \rt} [1 - N(\rt)],
\end{equation}
which is directly proportional to the unintegrated gluon distribution and to the leading order quark-target cross section.
After the BK evolution, the 2D Fourier transform remains positive up to larger values of transverse momenta. However, given the rather steep initial dipole amplitude (large $\g$), especially in the Balitsky+smallest dipole setup, a very long evolution is necessary to render the Fourier transform positive definite in the phenomenologically relevant transverse momentum range. 

We have also considered the ``ResumBK'' equation (following the terminology of Ref.~\cite{Beuf:2020dxl}), which resums single transverse logarithms~\cite{Iancu:2015joa} on top of the double logs~\cite{Iancu:2015vea} included also in the KCBK equation. Using the same MV$^\g$ parametrization, the setup produced fits that were not able to describe the HERA data, the best-fit parametrization resulting in global $\chi^2/\mathrm{d.o.f.}\sim1.9$ (and $\chi^2/\mathrm{d.o.f.}\sim5.5$ when comparing to the charm data only). This setup also required an anomalous dimensions $\g \gtrsim 4$ at the initial condition, which would also result in negative dipole amplitudes at small-$r$ as a result of the evolution\footnote{In Ref.~\cite{Beuf:2020dxl} this problem was avoided by enforcing $N(r)\ge 0$.}. 
We also considered the MV$^e$ initial condition parametrization from Ref.~\cite{Lappi:2013zma} with the ResumBK evolution, which corresponds to Eq.~\eqref{eq:bk-ic} with fixed $\g = 1$, but it did not result in better fit quality as well (total $\chi^2/\mathrm{d.o.f.}\gtrsim 2.5$).
For these reasons, in this work we only report results obtained with the KCBK evolution.  In the future, we plan to couple these single and double log resummations included in the ResumBK equation, and the full next-to-leading order BK equation~\cite{Balitsky:2008zza}, into the global analysis setup developed in this work. 

\section{Conclusions and Outlook}
\label{sec:conclusions}

We have constrained the model parameters describing the non-perturbative initial condition of the small-$x$ BK equation at next-to-leading order accuracy through a global analysis including the total cross section and charm production data from  HERA.
These datasets are found to  provide tight constraints for the model parameters, as demonstrated in the posterior distributions extracted using Bayesian inference and shown in Figs.~\ref{fig:kcbk_balsd_corner} and~\ref{fig:kcbk_pd_corner}. 
Compared to the previous NLO fits~\cite{Beuf:2020dxl,Hanninen:2022gje} in the dipole picture, this work corresponds to the first global analysis with heavy quark contribution included,  provides uncertainty estimates for the non-perturbative parameters and accounts for the correlated systematic uncertainties in the HERA data.
The resulting posterior distributions can be used to propagate uncertainties in the initial condition of the BK equation to all CGC calculations at next-to-leading order accuracy, including also heavy quark contributions. Parametrization vectors sampled from the posterior distributions, as well as BK evolved dipole amplitudes, are available at~\cite{samples_zenodo}.

In this work we combined the NLO impact factors to the approximated version of the NLO BK evolution equation. Results with two different running coupling prescriptions, parent dipole and Balitsky+smallest dipole, are reported. The Balitsky+smallest dipole prescription is physically better motivated and it results in model parameters that are also favored by exclusive vector meson production data. On the other hand, using both running coupling setups there is some tension between the total cross section and the charm production data, and this tension is stronger when the Balitsky+smallest dipole running coupling scheme is used.

An analysis using the full NLO BK equation is the natural next step for this work. This will include, in the BK equation, contributions not enhanced by large transverse logarithms captured by resummation prescriptions, and additionally a resummation of single transverse logarithms not included in the KCBK equation used in this work. This can be achieved following the setup developed in Ref.~\cite{Lappi:2016fmu}. Furthermore, it will also be important to repeat the analysis using the recently derived BK equation formulated in terms of the target rapidity evolution~\cite{Ducloue:2019ezk}, and to include other observables such as diffractive cross section~\cite{Beuf:2024msh} to the global analysis. 
 
\begin{acknowledgments}
C.C and H.M are supported by the Research Council of Finland, the Centre of Excellence in Quark Matter, and projects 338263, 346567 and 359902, and by the European Research Council (ERC, grant agreements  No. ERC-2023-101123801 GlueSatLight and ERC-2018-ADG-835105 YoctoLHC).
C.C acknowledges the support of the Vilho, Yrjö and Kalle Väisälä Foundation.
H.H is supported by the Research Council of Finland (Flagship of Advanced Mathematics for Sensing Imaging and Modelling grant 359208; Centre of Excellence of Inverse Modelling and Imaging grant 353092), and the Vilho, Yrjö and Kalle Väisälä Foundation.
Computing resources from CSC – IT Center for Science in Espoo, Finland and the Finnish Grid and Cloud Infrastructure (persistent identifier \texttt{urn:nbn:fi:research-infras-2016072533}) were used in this work.
The content of this article does not reflect the official opinion of the European Union and responsibility for the information and views expressed therein lies entirely with the authors. 

\end{acknowledgments}

\bibliographystyle{JHEP-2modlong.bst}
\bibliography{refs}

\clearpage 

\end{document}